\documentclass{article}
\usepackage[T1]{fontenc}
\usepackage{amsmath, amssymb}
\usepackage{graphicx}
\usepackage{booktabs, array}
\usepackage[labelfont=bf]{caption}
\usepackage{enumitem}
\usepackage{float}
\usepackage{subcaption}
\usepackage[utf8]{inputenc}
\usepackage{xcolor}
\usepackage{siunitx}
\usepackage{soul}
\usepackage[margin=0.75in]{geometry}
\usepackage{placeins}
\usepackage{multirow}
\usepackage{doi}
\usepackage{hyperref}  

\usepackage[numbers,sort&compress]{natbib}

\hypersetup{
    colorlinks=true,
    linkcolor=blue,
    filecolor=magenta,      
    urlcolor=cyan,
    pdftitle={Thermal Transport in Superlattices},
    pdfauthor={A. Chatterjee et al.},
    citecolor=blue,  
}
\title{Beam-Offset Thermoreflectance with Bayesian Optimization to Measure the Anisotropic Thermal Properties of Semiconductor Superlattices}

\author{%
A. Chatterjee\textsuperscript{1,2},
N. Spitzer\textsuperscript{2},
T. Kruck\textsuperscript{2},
P. Song\textsuperscript{3},\\[4pt]
A. Ludwig\textsuperscript{2},
A. D. Wieck\textsuperscript{2},
J. Ordonez-Miranda\textsuperscript{4},
M. Pawlak\textsuperscript{1,*}
}

\date{December 2025}

\begin{document}
\maketitle

\begin{center}
\textsuperscript{1}Institute of Physics, Faculty of Physics, Astronomy and Informatics, Nicolaus Copernicus University in Toruń, Grudziądzka 5, 87-100 Torun, Poland

\textsuperscript{2}Chair of Applied Solid-State Physics, Experimental Physics VI, Ruhr-University Bochum, Universitaetsstrasse 150, D-44780 Bochum, Germany

\textsuperscript{3}College of Mechanical and Transportation Engineering, China University of Petroleum, Beijing, China

\textsuperscript{4}CNRS, Institut des Nanosciences de Paris, University of Sorbonne, Paris, France

\vspace{0.3cm}
*Corresponding author: \texttt{mpawlak@fizyka.umk.pl}
\end{center}

\vspace{1cm}

\begin{abstract}
The directional nature of heat conduction in semiconductor superlattices---marked by significant differences between in-plane and cross-plane pathways---poses substantial challenges for precise thermal property assessment. Conventional frequency-domain thermoreflectance (FDTR) techniques, while proficient at evaluating cross-plane thermal conductivity, suffer from restricted capability in resolving in-plane transport due to inherent phase-delay constraints and inadequate lateral resolution. In this investigation, we establish a non-contact beam-offset FDTR (BO-FDTR) approach that concurrently determines both directional thermal conductivities within layered semiconductor architectures. Our methodology implements spatial separation between excitation and detection beams while utilizing coupled normalized amplitude and phase responses as analytical inputs, thereby improving discrimination between anisotropic thermal parameters. We combine this experimental configuration with a Bayesian optimization scheme incorporating Gaussian Process Regression (BO-GPR) to reduce estimation inaccuracies, attaining measurement uncertainties under \SIrange{1}{2}{\percent} at \SI{95}{\percent} confidence intervals. This technique demonstrates particular efficacy for intricate multilayer nanostructures, furnishing a structured protocol for superlattice thermal evaluation. Experimental characterization of an AlAs/GaAs superlattice (period thickness \SI{52}{\nano\meter}) delivers thermal conductivity values of \SI{14.7}{\watt\per\meter\per\kelvin} (cross-plane) and \SI{37.4}{\watt\per\meter\per\kelvin} (in-plane). Our findings indicate that integrating frequency sweeps with varied beam offset locations yields superior measurement precision, exceeding conventional single-variable methods and confirming thermal assessment validity across both geometric arrangements.
\end{abstract}

\section{INTRODUCTION}

Determining direction-specific thermal transport represents a critical frontier in modern materials science, with substantial implications for energy harvesting technologies \cite{Zhao2014} and quantum material engineering \cite{Chu2010}. Synthetic nanostructures, including semiconductor superlattices, manifest pronounced thermal directional dependence—the lateral thermal conductivity (\(k_{\parallel}\)) routinely deviates significantly from through-thickness values (\(k_{\perp}\)) through interface-governed phonon dynamics \cite{Medvedev2015}. These anisotropic thermal attributes fundamentally control operational characteristics in high-power electronic components \cite{Li2014}, heat dissipation assemblies \cite{Wieg2015}, and advanced thermoelectric converters \cite{Yan2010}. Established thermal characterization protocols confront inherent constraints when examining anisotropic material platforms. Frequency-domain thermoreflectance (FDTR), while proficient at evaluating cross-plane thermal transport in layered constructs \cite{Cahill2004}, demonstrates negligible sensitivity to transverse heat propagation in conventional aligned arrangements due to swift thermal signal dissipation \cite{Zhu1998}. This directional preference generates problematic parameter interdependence and considerable measurement variability for anisotropic specimens \cite{Pawlak2020}. Comparable complications arise for well-established techniques such as the 3\(\omega\) method: despite its effectiveness for isotropic materials \cite{Borca2001}, its implementation for anisotropic substrates and thin films introduces analytical complexities \cite{Ramu2012,Tong2006}.

Spatially separated experimental geometries alleviate these constraints by physically decoupling thermal excitation zones from detection regions, thereby intensifying responsiveness to in-plane thermal transport mechanisms \cite{Yang2013}. These offset beam methodologies, functional in both temporal \cite{Feser2012} and frequency-modulated \cite{Xu2023} operational modes, reconfigure thermal analysis into a sophisticated multivariate inversion challenge. Acquired signals transform into convoluted functions of several interconnected variables, with conventional optimization approaches often converging to local optima or generating unphysical uncertainty estimates \cite{Diessner2022}. Contemporary methodological innovations have enriched the analytical repertoire for anisotropic material investigation. Integrated photothermal infrared radiometry and thermoreflectance frameworks provide complementary perspectives on nanoscale thermal characteristics \cite{Chatterjee2024b}, while multiprobe thermal transport techniques offer alternative interrogation strategies for nanostructured systems \cite{Kim2015}. These developments highlight a growing consensus: thorough thermal property determination necessitates unified approaches addressing both experimental detection capability and computational deconvolution requirements.

The characterization imperative extends considerably beyond conventional semiconductor platforms to encompass varied material categories exhibiting thermal anisotropy. Exceptional lateral thermal conductivity documented in aluminum nitride thin films \cite{Hoque2021} and unusual directional transport within amorphous silicon nanostructures \cite{Kwon2017} illustrate the spectrum of phenomena requiring accurate measurement. Similarly, anisotropic thermal conduction in layered black phosphorus architectures \cite{Lee2015} and complex transport phenomena within reflective Mo/Si multilayers \cite{Medvedev2015} emphasize the necessity for methodologies capable of isolating directional thermal contributions. Addressing these challenges, we introduce a unified methodological framework merging beam-offset frequency-domain thermoreflectance (BO-FDTR) with Bayesian optimization employing Gaussian Process Regression (BO-GPR). Our strategy systematically interrogates thermal response across dual experimental coordinates—modulation frequency and spatial displacement—producing a comprehensive dataset where signatures of in-plane and cross-plane thermal transport become computationally distinguishable \cite{Pawlak2021}. This approach builds upon established photothermal radiometry principles \cite{Pawlak2018} while integrating contemporary insights regarding measurement precision optimization \cite{Chatterjee2024}.

The analytical core implements Bayesian statistical inference \cite{Gelman2013}, guided by statistical inversion concepts \cite{Bui2006}. By processing complete frequency-displacement datasets simultaneously via Gaussian Process modeling \cite{Rasmussen2006}, our methodology naturally disentangles correlated thermal parameters while providing rigorous uncertainty quantification through posterior probability distributions. This statistical architecture, supported by information-theoretic foundations \cite{MacKay2003}, enables robust parameter estimation resilient to experimental noise and model uncertainties. The computational forward model incorporates three-dimensional anisotropic heat conduction formulations that accurately represent suppressed lateral thermal diffusion in anisotropic media. This mathematical framework extends established thermoreflectance analytical techniques \cite{Schmidt2010} while specifically addressing anisotropic system complexities. The methodology's inherent flexibility permits application across material categories, from luminescent rare-earth-doped aluminum nitride ceramics \cite{Wieg2012} to intricate moiré superlattice architectures \cite{Ran2025}.

Experimental implementation employs precision optomechanical control for beam positioning \cite{Yang2014} and advanced phase-sensitive detection for signal integrity maintenance \cite{Chatterjee2024}. This integrated platform facilitates simultaneous measurement of amplitude and phase responses with required precision to resolve subtle anisotropic signatures. The approach constitutes a substantive advance beyond conventional univariate thermal characterization paradigms. This investigation demonstrates our comprehensive framework through determination of anisotropic thermal conductivity in an AlAs/GaAs superlattice system. Our methodology successfully extracts precise, physically consistent values for both \(k_{\parallel}\) and \(k_{\perp}\) while rigorously quantifying measurement uncertainties through Bayesian inference. This work establishes a broadly applicable protocol for anisotropic thermal property determination, equipping researchers with an advanced toolkit for thermal design and performance optimization of next-generation nanoscale material systems and functional devices.

\section{Material Preparation Protocol}

An undoped, \qty{76}{\milli\meter} diameter (001)-oriented GaAs wafer served as the epitaxial substrate. Growth initiated with thermal oxide desorption followed by deposition of a \qty{150}{\nano\meter} GaAs buffer layer at \qty{600}{\degreeCelsius}. Subsequently, an AlAs/GaAs superlattice consisting of 10 periods was grown, with each period comprising alternating \qty{26}{\nano\meter} layers of AlAs and GaAs, terminating with a GaAs layer. Post-growth, the wafer was cleaved into \qty{10}{\milli\meter} $\times$ \qty{12}{\milli\meter} pieces. A \qty{50}{\nano\meter} Au transducer layer was deposited via tungsten-boat evaporation in a high-vacuum chamber ($<$ \qty{e-6}{\milli\bar}) \cite{Pawlak2020}

\begin{table}[h!]
\centering
\caption{Structural parameters of the AlAs/GaAs superlattice}
\label{tab:thickness_specs}
\begin{tabular}{lc}
\toprule
\textbf{Parameter} & \textbf{Value} \\
\midrule
\textit{Transducer} & \\
\quad Au thickness & \SI{0.05}{\micro\meter} \\
\midrule
\textit{Superlattice} & \\
\quad Period (AlAs/GaAs) & 10 × \SI{26}{\nano\meter}/\SI{26}{\nano\meter} \\
\quad Period thickness & \SI{52}{\nano\meter} \\
\quad Total SL thickness & \SI{0.520}{\micro\meter} \\
\quad (ending with GaAs) & \\
\midrule
\textit{Substrate} & \\
\quad Material & GaAs (001) \\
\quad Doping & undoped \\
\quad Thickness & \SI{520}{\micro\meter} \\
\bottomrule
\end{tabular}
\end{table}

Optical reflectance confirmed the total superlattice thickness as \qty{520}{\nano\meter}, consistent with the target design \cite{Pawlak2020}.

\begin{figure}[H]
    \centering
    \includegraphics[width=0.95\linewidth]{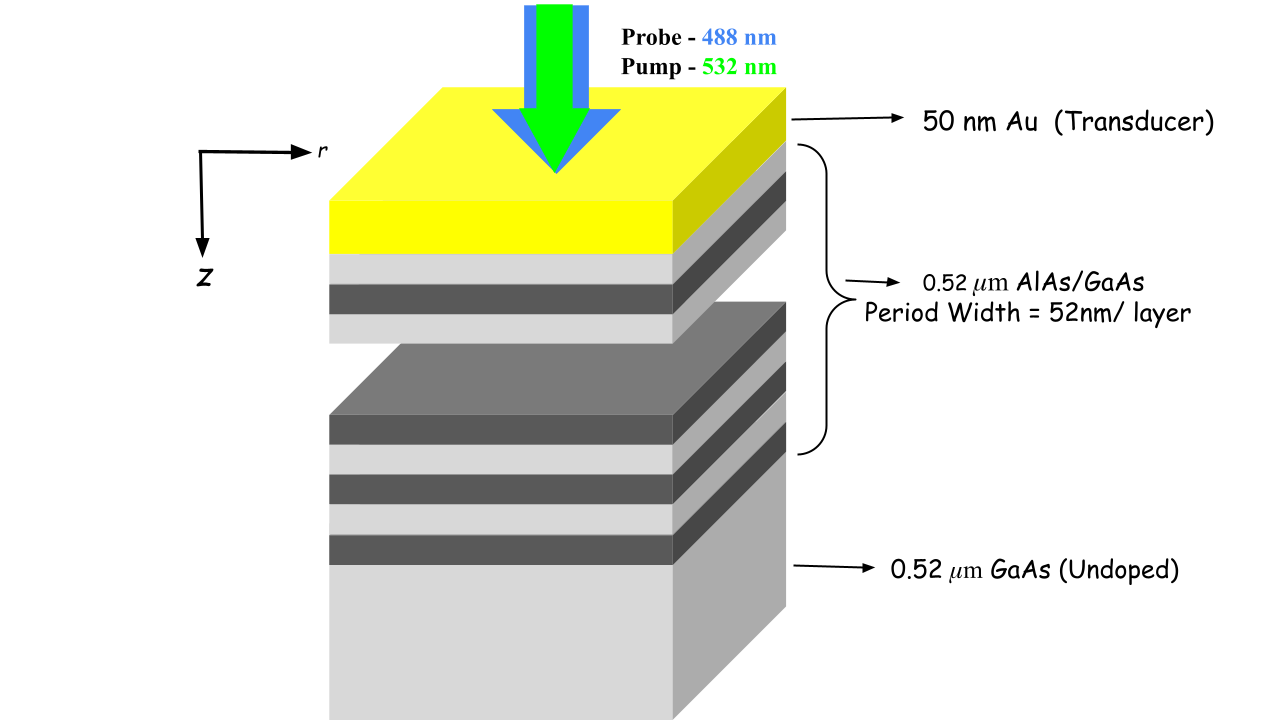}
    \caption{Schematic of the sample structure and experimental configuration for beam-offset thermoreflectance measurements. The sample consists of a gold transducer layer deposited on a GaAs/AlAs superlattice with total thickness $d_1 = \SI{0.52}{\micro\meter}$ and an undoped GaAs substrate with thickness $d_2 = \SI{520}{\micro\meter}$. The superlattice period is $\Lambda = \SI{52}{\nano\meter}$, corresponding to 10 bilayers. A modulated pump laser beam at wavelength $\lambda = \SI{532}{\nano\meter}$ with Gaussian spatial profile generates thermal waves via the intensity distribution. The beam offset $\Delta r$ between pump and probe beams enables depth-resolved thermal characterization of the multilayer structure.}
    \label{fig:Figure1}
\end{figure}

\section{Mathematical Model Fitting}

In this section, the 3D model for BO-FDTR is introduced. For notation purposes in mathematical modeling, the in-plane component is denoted by $k_{\parallel}$ and the cross-plane by $k_{\perp}$ (notations used throughout the paper). In the Hankel space ($\lambda$ = Hankel variable) both temperature fluctuation and heat flux are functions of the layer thickness. The thermal response detected by the probe beam (in a 2-layer system) has been used as a reference signal for normalization of both phase and amplitude \cite{Pawlak2020} in the n-layer system. Following the matrix formalism established in prior work \cite{Pawlak2020}, the expression for thermal responses at each modulation frequency $f$ and $\Delta r$ \textit{i.e.} Beam offset distance between pump and probe lasers is discussed in the following sections. The anisotropic thermal parameter optimization framework (Bayesian) requires a physical model that accurately predicts the thermal responses given the parameters of the \textit{n}-layer material. This section further develops the multi-layer 3D anisotropic heat diffusion model that serves as the forward model in the Bayesian inference process.

\subsection*{1-layer system}
We formulate a three-dimensional thermal transport model for axisymmetric heat diffusion in cylindrical coordinates, where angular independence simplifies the analysis of single-layer systems as {\cite{Pawlak2020, Yang2013}:

\begin{equation}
k_{\parallel} \left( \frac{\partial^2 T}{\partial r^2} + \frac{1}{r} \frac{\partial T}{\partial r} \right) + k_{\perp} \frac{\partial^2 T}{\partial z^2} = C\frac{\partial T}{\partial t}
\label{eq:1}
\end{equation}

where \(r\) represents the radial coordinates contribute towards the radial thermal diffusion, while \(z\) represents the depth direction constituting for the vertical diffusion respectively. Here,\(C\) is the cylindrically symmetrical heating source which determines the contribution of each part. A strategic solution to Eq.(~\eqref{eq:1} is -

\begin{equation}
T(r, z, t) = T_0(r,z) + \Re\left[ \Theta(r,z) e^{-j\omega t} \right]
\label{eq:2}
\end{equation}

where \( T(r,z,t) \) is the total temperature field, decomposed into a steady-state component \( T_0(r,z) \) (steady-state temperature) and a modulated component \( \Theta(r,z) \) (complex amplitude of the oscillatory temperature variation).  Substituting Eq.~\eqref{eq:2} into Eq.~\eqref{eq:1} and simplifying: 
\begin{equation}
\left( \frac{\partial^2 T}{\partial r^2} + \frac{1}{r} \frac{\partial T}{\partial r}\right) + \frac{k_\perp}{k_\parallel} \frac{\partial^2 T}{\partial z^2} = \frac{C} {k_\parallel} \frac{\partial T}{\partial t}
\label{eq:3}
\end{equation}
where \(\alpha = \frac{k_{\parallel}}{C}\) is the thermal diffusivity in the parallel direction.
Applying a Hankel transform to the radial component of Eq.~\eqref{eq:1} yields:
\begin{equation}
-\lambda^2 \Theta + \frac{k_{\perp}}{k_{\parallel}}\frac{\partial^2 \Theta}{\partial z^2} = \frac{i\omega}{\alpha} \Theta \quad \Rightarrow \quad \frac{\partial^2 \Theta}{\partial z^2} - \sigma^2 \Theta = 0
\label{eq:4}
\end{equation}
To facilitate solution, we introduce the complex propagation parameter:
\begin{equation}
\sigma^2 = \frac{k_{\parallel}}{k_{\perp}} \left( \lambda^2 + \frac{i\omega}{\alpha} \right), \quad
\omega = 2\pi f, \quad
\eta = \frac{k_\perp}{k_\parallel}
\label{eq:5}
\end{equation}
Here $f$ represents the pump modulation frequency, with complete variable definitions provided in Tables~\ref{tab:thickness_specs} and~\ref{tab:abbreviations}. Solving Eq.~\eqref{eq:4} with the parameters defined in Eq.~\eqref{eq:5} produces:
\begin{equation}
\Theta(z) = A' e^{\sigma z} + B' e^{-\sigma z}
\label{eq:6}
\end{equation}
here \(A'\) and \(B'\) are the dimensionality constants. Similarly on the other hand, the heat flux \( \phi(z) \),
\begin{equation}
\Phi(z) = -k_{\perp} \frac{\partial \Theta}{\partial z} = -k_{\perp} \sigma \left( A' e^{\sigma z} - B'^{-\sigma z} \right)
\label{eq:7}
\end{equation}

Hence at the surface (\(z=0\)) from Eq.~\eqref{eq:6} and Eq.\eqref{eq:7}
\begin{equation}
\frac{B'}{A'} = \frac{\Theta(0) + \frac{\Phi(0)}{k_{\perp} \sigma}}{\Theta(0) - \frac{\Phi(0)}{k_{\perp} \sigma}}
\label{eq:8}
\end{equation}

Replacing the ratio $\frac{B'}{A'}$ for the representation of $\Theta(z)$ and $\Phi(z)$ in terms of $A'$ and $B'$ along cross-plane geometry, we obtain:

\begin{equation}
\Phi(0) = k_{\perp} \sigma (B' - A')
\label{eq:9}
\end{equation}

which in turn allows to find the heat flux and the temperature field at \(z\) = 0 at the surface. Hence the expression for the temperature fluctuations at depth \(z\) is a function of the heat flux and temperature fluctuations which is given by-

\begin{equation}
\Theta(z) = \Theta(0) \cosh(\sigma z) - \frac{\Phi(0)}{k_{\perp} \sigma} \sinh(\sigma z)
\label{eq:10}
\end{equation}

\begin{equation}
\Phi(z) = -k_{\perp} \sigma \left[ \Theta(0) \sinh(\sigma z) - \frac{\Phi(0)}{k_{\perp} \sigma} \cosh(\sigma z) \right]
\label{eq:11}
\end{equation}
 Solving by replacing ~\eqref{eq:10} and ~\eqref{eq:11} as  contribution of single layer $(z = l)$ in matrix representation respectively-

 \begin{equation}
\begin{bmatrix}
\Theta^{(l)} \\
\Phi^{(l)}
\end{bmatrix}
=
\begin{bmatrix}
A & B \\
C & A
\end{bmatrix}
\begin{bmatrix}
\Theta^{(0)} \\
\Phi^{(0)}
\end{bmatrix} =
\begin{bmatrix}
\cosh(\sigma l) & \dfrac{\sinh(\sigma l)}{k_\perp \sigma} \\
- k_\perp \sigma \sinh(\sigma l) & \cosh(\sigma l)
\end{bmatrix}
\begin{bmatrix}
\Theta^{(0)} \\
\Phi^{(0)}
\end{bmatrix} 
\label{eq:12}
\end{equation}
 where,
\begin{equation}
\begin{aligned}
A &= \cosh(\sigma l) \\
B &= \dfrac{-\sinh(\sigma l)}{k_\perp \sigma} \\
C &= -k_{\perp}  \sigma \sinh(\sigma l)
\end{aligned}
\label{eq:13}
\end{equation}
Similar for the thermal transport for the three-layer system it is assumed that the contribution of each interface to the temperature fluctuation is detected by the probe beam (same as Eq.~\eqref{eq:11} and~\eqref{eq:12}).For a semi-infinite substrate, the boundary condition is considered (layer~3) at any interface can be deduced by the boundary condition where the heat flux is continuous. This allows us to write:

\begin{equation}
\Phi(l_+) = \Phi(l_-) = \frac{\Theta(l_-) - \Theta(l_+)}{R}.
\label{eq:14}
\end{equation}

At the interface, Matrix contribution of thermal boundary resistance $R_{\text{th}}$ determining the continuity of the temperature field from Eqs.~\eqref{eq:14} and~\eqref{eq:12}:

\begin{equation}
\begin{bmatrix}
\theta(l_+) \\
\phi(l_+)
\end{bmatrix}
=
\begin{bmatrix}
1 & -R \\
0 & 1
\end{bmatrix}
\begin{bmatrix}
\theta(l_-) \\
\phi(l_-)
\end{bmatrix}
\label{eq:15}
\end{equation}

where $R = -\dfrac{D}{C}$ from Eq.~\eqref{eq:14}.

\subsection*{3-Layer system}

Similar to thermal transport in a three-layer system, it is assumed that the contribution of each interface to the temperature fluctuation is detected by the probe beam (same as Eqs.~\eqref{eq:10} and~\eqref{eq:11}). For a semi-infinite substrate, the following boundary condition is considered (layer 3):

\begin{equation}
\hat{\Phi}_3 = 0.
\label{eq:16}
\end{equation}

The temperature and flux fields at the top of each layer ($z_n = l_n$) are:

\begin{equation}
\begin{cases}
\hat{\Theta}_n = \hat{\Theta}(z_1+z_2+z_3) \\
\hat{\Phi}_n = \hat{\Phi}(z_1+z_2+z_3)
\end{cases}
\label{eq:17}
\end{equation}

The contribution of the thermal boundary resistance at the interface between the $n$-th and $(n+1)$-th layer in matrix representation:
\begin{equation}
\mathbf{R}_{n,n+1} = 
\begin{bmatrix}
1 & -R_{n,n+1} \\
0 & 1
\end{bmatrix}.
\label{eq:18}
\end{equation}
The transfer matrix representing the contribution for each layer (from Eq:~\eqref{eq:13}) is:

\begin{equation}
\mathbf{M}_n =
\begin{bmatrix}
A'_n & B'_n \\
C'_n & A'_n
\end{bmatrix},
\quad
\begin{aligned}
A'_n &= \cosh(\sigma_n z_n) \\
B'_n &= \frac{\sinh(\sigma_n z_n)}{k_{\perp n} \sigma_n} \\
C'_n &= -k_{\perp n} \sigma_n \sinh(\sigma_n z_n)
\end{aligned}
\label{eq:19}
\end{equation}

Applying the boundary condition $\hat{\Phi}_3 = 0$ to Eqs.~\eqref{eq:10} and~\eqref{eq:11} gives:

\begin{equation}
0 = C\cdot \hat{\Theta}_0 + D \cdot \hat{\Phi}_0 \quad \Longrightarrow \quad \hat{\Theta}_0 = \hat{\Theta}(0) = -\frac{D}{C} \hat{\Phi}(0).
\label{eq:20}
\end{equation}

Hence for overall 3-layer superlattice nanostructure, TR response Eq:~\eqref{eq:17} can be represented in matrix formation as:
\begin{equation}
\begin{bmatrix}
\hat{\Theta}_3 \\
\hat{\Phi}_3
\end{bmatrix}
=
\begin{bmatrix}
A & B \\
C & D
\end{bmatrix}
\begin{bmatrix}
\hat{\Theta}(0) \\
\hat{\Phi}(0)
\end{bmatrix},
\quad
\begin{bmatrix}
A & B \\
C & D
\end{bmatrix}
= \mathbf{M}_3 \mathbf{R}_{23} \mathbf{M}_2 \mathbf{R}_{12} \mathbf{M}_1.
\label{eq:21}
\end{equation}
In case of the FDTR pmup-probe analysis, the frequency-modulated Gaussian pump beam intensity profile can be represented as:
\begin{equation}
I(r,t) = \frac{2Q}{\pi \sigma_0^2} \exp\left[-2\left(\frac{r}{a}\right)^2\right] \big(1+ \cos(\omega t)\big)
\label{eq:22}
\end{equation}
where $Q$ is the total power, $\sigma_0$ is the beam waist radius ($1/e^2$ intensity radius), 
$a = \sigma_0/\sqrt{2}$ is the $1/e$ intensity radius, and $\omega$ is the pump modulation frequency. In FDTR analysis, the Hankel transform of the Gaussian intensity profile converts the radial heat distribution into the frequency domain, providing the analytical solution. The surface temperature response in transform space correlates with the experimentally measured thermoreflectance signal amplitude and Phase. Replacing Eq: ~\eqref{eq:22} we get the TR signal as:

\begin{equation}
\hat{\Phi}(0) = \mathcal{H}_0\left[ I(r) \right] = \frac{2Q}{\pi\sigma_0^2} \int_{0}^{\infty} e^{-(r/a)^2} \, J_0(\lambda r) \, r \, dr = \frac{Q}{2\pi} \exp\left(-\frac{\lambda^2 \sigma_0^2}{8}\right)
\label{eq:23}
\end{equation}
where $J_0(\lambda r)$ is the Bessel function of the first kind of order zero (Hankel transform kernel).

For the temperature response field, we must account for the contribution of individual layers and their respective interface boundary resistances, leading to:

\begin{equation}
\begin{bmatrix}
A & B \\
C & D
\end{bmatrix}
= \mathbf{M}_3 \mathbf{R}_{23} \mathbf{M}_2 \mathbf{R}_{12} \mathbf{M}_1.
\label{eq:24}
\end{equation}

The components breaking down indicate the contribution of individual layers and interfaces ($\mathbf{M}_3$, $\mathbf{M}_2$, $\mathbf{M}_1$, similar to Eq.~\eqref{eq:21}):

\begin{equation}
\mathbf{M}_n = 
\begin{bmatrix}
A_n & B_n \\
C_n & A_n
\end{bmatrix}.
\label{eq:25}
\end{equation}

The values of $A_n$, $B_n$, $C_n$ are calculated for each layer using Eq.~\eqref{eq:13}. Further simplification of the semi-infinite thermally isotropic substrate from the following boundary condition:

\begin{equation}
l_3 \to \infty,\quad |\sigma_3| l_3 \gg 1,\quad k_{13} = k_1.
\label{eq:26}
\end{equation}

The contribution from each layer and its subsequent interface can be represented in matrix notation as:
\begin{equation}
\underbrace{
\begin{bmatrix}
A & B \\
C & D
\end{bmatrix}}_{\mathbf{M}_{\mathrm{total}}}
=
\underbrace{
\begin{bmatrix}
A_3 & B_3 \\
C_3 & D_3
\end{bmatrix}}_{\mathbf{M}_3}
\underbrace{
\begin{bmatrix}
\bar{\alpha} & \bar{\beta} \\
\bar{\gamma} & \bar{\delta}
\end{bmatrix}}_{\widetilde{\mathbf{M}}},
\label{eq:27}
\end{equation}

where the intermediate matrix $\widetilde{\mathbf{M}}$ accounts for all layers and interfaces:

\begin{equation}
\widetilde{\mathbf{M}} = \mathbf{M}_3
\underbrace{
\begin{bmatrix}
1 & -R_{32} \\
0 & 1
\end{bmatrix}}_{\mathbf{R}_{32}}
\mathbf{M}_2
\underbrace{
\begin{bmatrix}
1 & -R_{21} \\
0 & 1
\end{bmatrix}}_{\mathbf{R}_{21}}
\mathbf{M}_1.
\label{eq:28}
\end{equation}

Here, $\mathbf{M}_i$ ($i=1,2,3$) from Eq.~\eqref{eq:21} is the $2 \times 2$ transfer matrix of layer $i$, and $\mathbf{R}_{ij}$ represents the $2 \times 2$ matrix for thermal resistance $R_{ij}$ from Eq.~\eqref{eq:18} at the interface between layers $i$ and $j$.

For isotropic substrates: $k_\perp = k_\parallel$ (GaAs substrate in our case), the boundary condition can help us solve for the independent parameters in each layer (from Eqs.~\eqref{eq:10},~\eqref{eq:11} in~\eqref{eq:25}):

\begin{align}
C &= \bar{\alpha} C_3 + \bar{\gamma} A_3,\label{eq:29} \\
D &= \bar{\beta} C_3 + \bar{\delta} A_3. \label{eq:30}
\end{align}

Substituting the value of $\frac{D}{C}$ and boundary condition Eq:~\eqref{eq:26}, Eqs.~\eqref{eq:18}, in the Eqs. ~\eqref{eq:10} and ~\eqref{eq:11}, we obtain:

\begin{equation}
\frac{D}{C} = \frac{\bar{\delta} - \sigma_3 k_3 \bar{\beta}}{\bar{\gamma} - \sigma_3 k_3 \bar{\alpha}}
\label{eq:31}
\end{equation}

where the notations ($n_{th} $ layer)  are:

\begin{equation}
\sigma_{1,3}^2 = \eta^2 \left( \lambda^2 + \frac{i\omega}{\alpha_{1,3}} \right), \quad
\sigma_2^2 = \eta^2 \left( \lambda^2 + \frac{i\omega}{\alpha_{1,2}} \right), \quad
\eta = \frac{k_\perp}{k_\parallel}.
\label{eq:32}
\end{equation}

\begin{align}
\bar{\alpha} &= A_2 (B_1 - R_{21} A_1) + B_2 A_1 - R_{32} [C_2 (A_2 - R_{21} A_1) + C_1 A_2], \nonumber \\
\bar{\beta} &= A_2 (B_1 - R_{21} A_1) + B_2 A_1 - R_{32} [C_2 (B_1 - R_{21} A_1) + A_1 A_2], \nonumber \\
\bar{\gamma} &= C_2 (A_2 - R_{21} A_1) + C_1 A_2, \nonumber \\
\bar{\delta} &= C_2 (B_1 - R_{21} A_1) + A_2 A_1.
\label{eq:33}
\end{align}

Finally, replacing Eq:~\eqref{eq:30}, Eq:~\eqref{eq:31},Eq:~\eqref{eq:32}, the surface temperature response in Hankel space is:

\begin{equation}
\hat{\Theta}(0) = -\frac{Q}{2\pi} \frac{D}{C} \exp\left(-\frac{\lambda^2 \sigma_0^2}{8}\right)
\label{eq:34}
\end{equation}

In real space for $\Delta r=0$ , applying the inverse Hankel transform:

\begin{equation}
\Theta(r,\omega) = -\frac{Q}{2\pi} \int_{0}^{\infty} \frac{D}{C} \exp\left(-\frac{\lambda^2 \sigma_0^2}{8}\right) J_0(\lambda r) \lambda \, d\lambda
\label{eq:35}
\end{equation}
where $J_0(0) = 1$ was used.

This is the "frequency-dependent" temperature response at the zero-offset position. For pump-probe measurements with added beam offset \(\Delta r\) and probe, pump beam waist radius \(\sigma_p\) , \(\sigma_0\), the measured signal is deduced from the boundary condition in Eq.~\eqref{eq:10}, Eq.~\eqref{eq:11} and Eq.~\eqref{eq:16}:

\begin{equation}
\Theta(\omega, \Delta r) = -\frac{Q}{2\pi} \int_{0}^{\infty} \frac{D}{C} \exp\left(-\frac{\lambda^2(\sigma_0^2 + \sigma_p^2)}{8}\right) J_0(\lambda \Delta r) \lambda \, d\lambda
\label{eq:36}
\end{equation}

We optimize the parameters by substituting all the values from Eqs.~\eqref{eq:32} and \eqref{eq:33} in Eq.~\eqref{eq:36}.
This denotes the underlying emphasis on the fact that the extracted $k_{\parallel}$ and $k_{\perp}$ (represented in Table I by $k_{\parallel}$ and $k_{\perp}$, respectively, throughout the entire manuscript) were determined independently in our measurements yet are nearly identical (to better than \SI{2.4}{\percent} for the superlattice), as required for this superlattice structure. The data orientation in Fig.~2 is shown accordingly.

\begin{table}[H]
\centering
\caption{Abbreviations and Acronyms}
\label{tab:abbreviations}
\begin{tabular}{cl}
\toprule
\textbf{Symbol} & \textbf{Description} \\
\midrule
$R_{n,m}$ & Thermal Boundary Resistance (TBR) (between $n$-$m$ layers) \\
$k_\perp$ & Cross-plane thermal conductivity (of $m$ layers), note $k_{\perp 2} \neq k_2$ \\
$k_\parallel$ & In-plane thermal conductivity (of $m$ layers)\\
$\sigma_0$ & Pump Beam Waist Radius ($1/e^2$ intensity radius) \\
$\sigma_p$ & Probe Beam Waist Radius ($1/e^2$ intensity radius) \\
$a = \sigma_0/\sqrt{2}$ & pump intensity radius ($1/e$) \\
$n$ & Layer index \\
$\alpha$ & Thermal Diffusivity\\
$\lambda$ & Hankel variable \\
\bottomrule
\end{tabular}
\end{table}

This model, computed on a three-layer superlattice structure (comprising a Gold (Au) transducer deposited on the AlAs/GaAs superlattice structure and GaAs substrate, see Fig.~2), solves the mystery of pronounced lateral heat spreading. This result, when compared to previous experiments on GaAs substrate \cite{Chatterjee2024} with its higher thermal conductivity, shows much improvement in thermally anisotropic samples. Further improvement regarding the uncertainty could be realized by shifting to a lower thermal conductivity superlattice for the thermal boundary resistance analysis, which is essentially important when studying dependence on its anisotropy index.

\begin{figure}[H]
    \centering
    \subfloat[Normalized Phase\label{fig:phase_comparison}]{
        \includegraphics[width=0.98\textwidth]{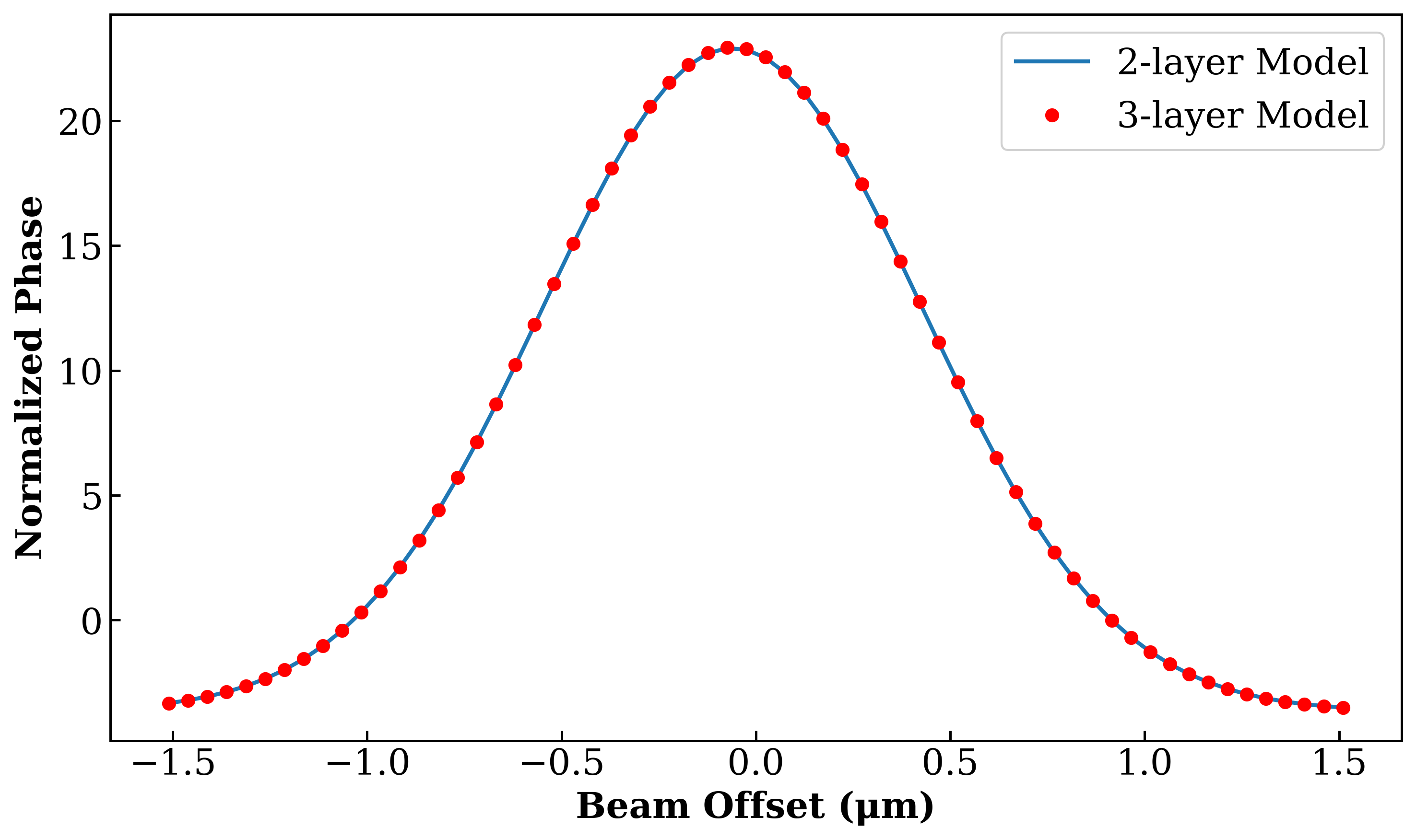}
    }
    \hfill
    \subfloat[Normalized Amplitude\label{fig:amplitude_comparison}]{
        \includegraphics[width=0.98\textwidth]{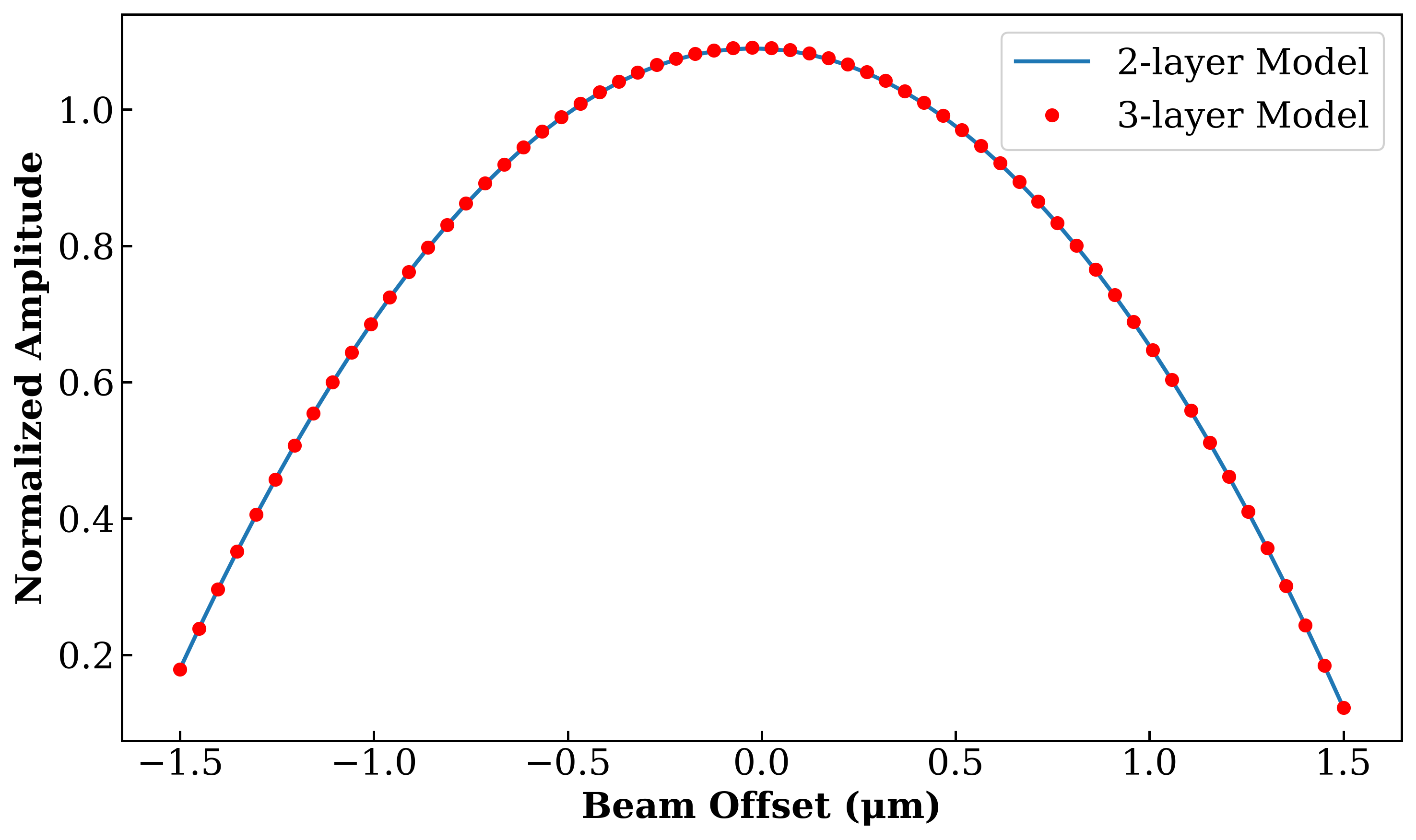}
    }
    \caption{ Comparative study of Normalized Amplitude and Phase between 2-layer model ($k = 2~\text{Wm}^{-1}\text{K}^{-1}$, $\alpha = 10^{-6}~\text{m}^{2}\text{s}^{-1}$) vs 3-layer model, $k = 2~\text{Wm}^{-1}\text{K}^{-1}$, $\alpha = 10^{-6}~\text{m}^{2}\text{s}^{-1}$ from \cite{Pawlak2020}}
    \label{fig:2}
\end{figure}

The derived expression for $\Theta(\omega, \Delta r)$ in Eq.~\eqref{eq:36} provides the forward model needed for Bayesian parameter estimation. By comparing predicted responses from this model with experimental measurements across multiple $(\Delta r, f)$ combinations, the Bayesian framework (Bayesian Optimisation) simultaneously extracts $k_{\parallel}$, $k_{\perp}$, and $R_{\text{th}}$ with quantified uncertainties.

\section{EXPERIMENTAL SETUP}\label{s:EXPERIMENTAL_SETUP}
The non-contact beam-offset FDTR system implements a dual-wavelength pump-probe configuration based on established thermoreflectance principles \cite{Cahill2004, Schmidt2010}. Thermal excitation is provided by a modulated 532 nm diode-pumped solid-state laser, while a continuous-wave 488 nm laser (Coherent Sapphire) serves as the probe. These optical paths are merged via a dichroic mirror and focused onto the sample surface using a 100× microscope objective with a numerical aperture of 0.9. Controlled spatial separation between the excitation and detection spots is achieved by rotating a beamsplitter mounted on a motorized rotational stage (Standa model 214377), enabling precise adjustment of the beam offset distance $\Delta r$. The thermoreflectance signal from the reflected probe beam is captured by a silicon photodiode and processed through a lock-in amplifier (Standard Instruments SR865A) referenced to the pump modulation frequency.

\begin{figure}[H]
    \centering
    \includegraphics[width=0.76\textwidth]{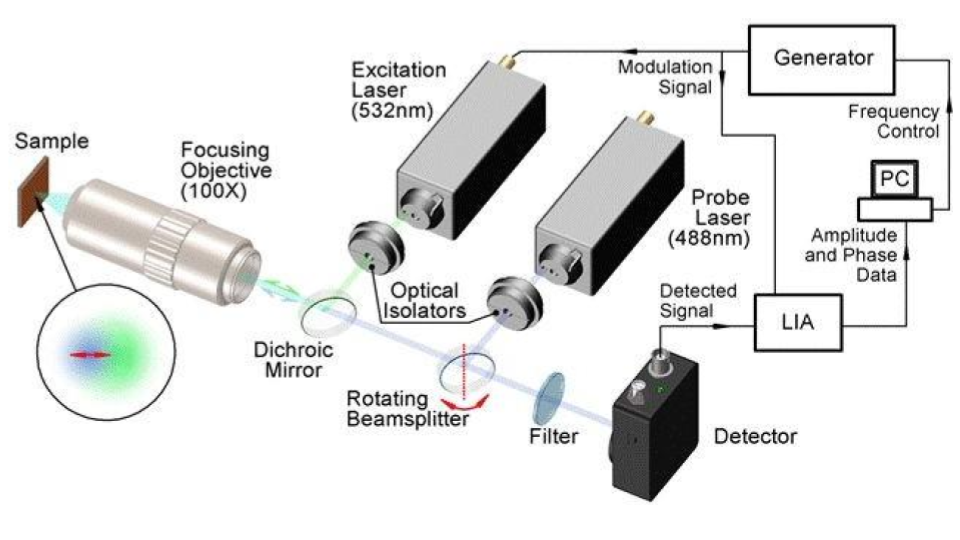}
    \caption{Schematic of the beam-offset frequency-domain thermoreflectance (BO-FDTR) setup. Key components include: AOM - Acousto-optic modulator (Crystal Technology Inc.) for pump beam modulation; M1 - steering mirror; LIA - lock-in amplifier for phase-sensitive detection; and a Standa 214377 rotational stage for controlled beam offset adjustment.}
    \label{fig:3}
\end{figure}

Beam offset control is implemented by rotating the beamsplitter using the precision rotational stage. The actuator provides a resolution of \SI{0.071}{\micro\meter} per step, enabling fine spatial control as illustrated in Fig.~\ref{fig:3}. The Standa rotational stage (Model 214377) is operated via an 8SMC5-USB driver interface programmed in LabVIEW, with an M27 aperture (\SI{30}{\milli\meter}) selected for optimal beam profiling for constant pump and probe spot size. For data analysis, a normalization procedure is applied using Eqs.~\eqref{eq:37} and \eqref{eq:38} \cite{Pawlak2020, Ran2025}. A reference sample consisting of a GaAs substrate with a \SI{50}{\nano\meter} Au transducer layer provides the baseline measurements. The selection of transducer material critically influences measurement quality, as the signal-to-noise ratio (SNR) and thermal response amplitude depend substantially on the material's thermoreflectance coefficient (CTR).

Normalization Constraints:
\begin{align}
\text{Normalized amplitude} \quad Amp(r,f) &= \frac{A_{\mathrm{n}}(r,f)}{A_{\mathrm{s}}(r,f)} \label{eq:37} \\
\text{Normalized phase} \quad Phase(r,f) &= Phase_{\mathrm{n}}(r,f) - Phase_{\mathrm{s}}(r,f) \label{eq:38}
\end{align}

\section{Sensitivity Analysis}

Sensitivity analysis quantifies how measurement signals respond to variations in thermal parameters. Following established thermoreflectance methodology \cite{Yang2013}, the sensitivity coefficient for beam-offset FDTR measurements at frequency $f_i$ is defined as:

\begin{equation}
\left[S_{x_j}\right]_{f_i} = \left[\frac{\partial \Phi}{\partial \ln x_j}\right]_{f_i} = \left[\frac{x_j}{\Phi} \frac{\partial \Phi}{\partial x_j}\right]_{f_i}
\label{eq:phase_sensitivity}
\end{equation}

Here, $\Phi$ represents either the phase ($\phi$) or amplitude ($\theta$) of the thermoreflectance signal, while $x_j$ denotes a thermal parameter such as $k_{\parallel}$ or $k_{\perp}$. The coefficient $\left[S_{x_j}\right]_{f_i}$ provides a dimensionless measure of sensitivity: a value of $\left[S_{x_j}\right]_{f_i} = 1$ indicates that a \SI{1}{\percent} variation in $x_j$ produces a \SI{0.01}{\degree} change in $\Phi$ (for phase signals). Parameters with $\left[S_{x_j}\right]_{f_i} > 1$ exhibit high sensitivity, where small changes in the parameter yield substantial signal variations, while values below unity indicate reduced sensitivity.
\begin{figure}[htbp]  
\centering
\includegraphics[width=0.95\textwidth]{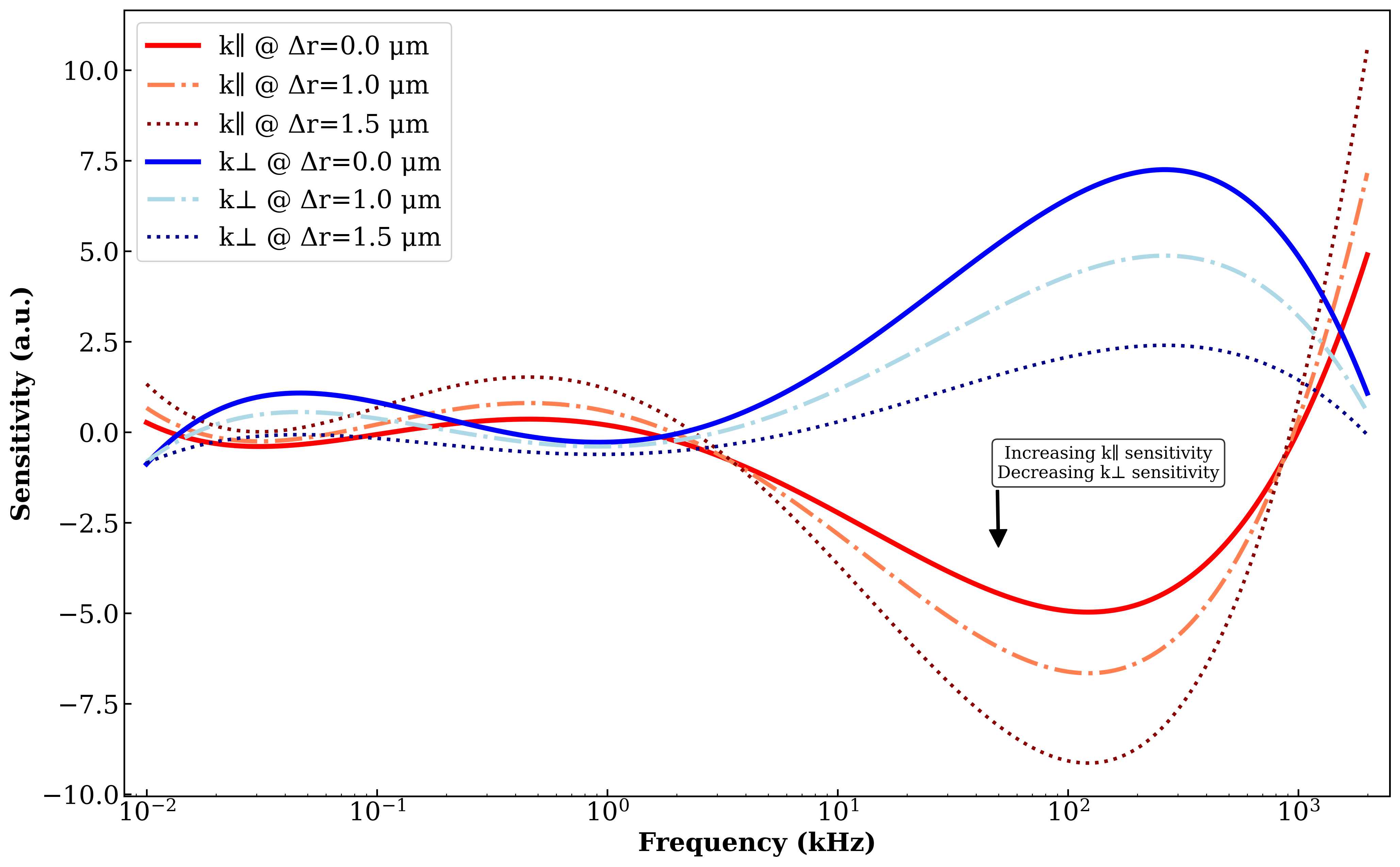}
\caption{Sensitivity of in-plane and cross-plane measurement to modulation frequency for AlAs/GaAs superlattice samples, showing TR phase response from analytical model at different offset position vs pump modulation frequency. The analysis covers the frequency range from \SI{1}{\kilo\hertz} to \SI{1.250}{\mega\hertz}.}
\label{fig:4}
\end{figure}

\begin{figure}[htbp]  
\centering
\includegraphics[width=1.0\textwidth]{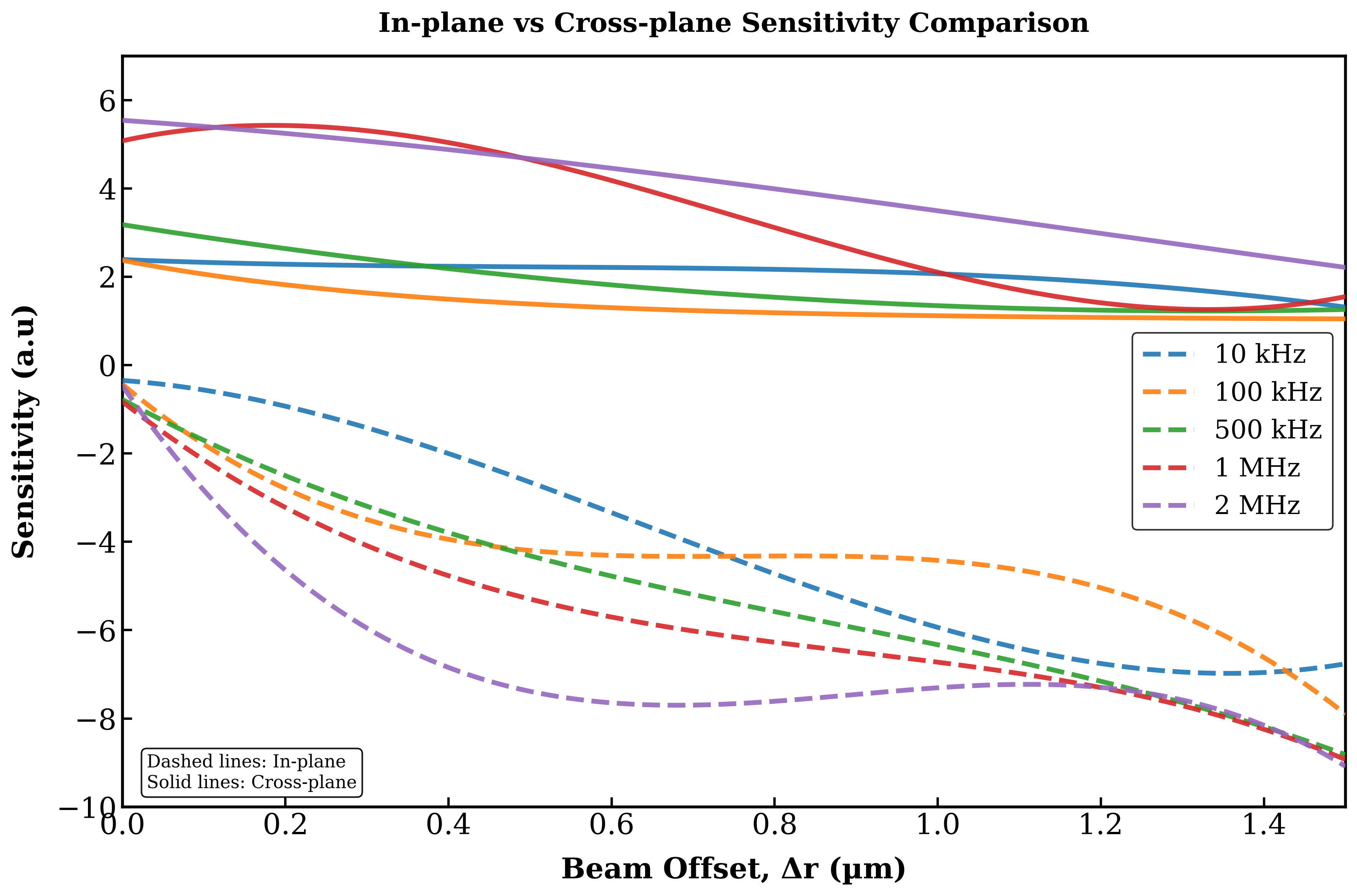}
\caption{Sensitivity of in-plane and cross-plane measurement to beam offsets $\Delta r = 0$ to $1.5~\mu$m.. for AlAs/GaAs superlattice samples, showing TR phase response from analytical model at different pump modulation frequency covers the frequency range from \SI{10}{\kilo\hertz} to \SI{2.0}{\mega\hertz}.}
\label{fig:sensitivity_freq2}
\end{figure}
The higher sensitivity of the beam offset mechanism as the scanning frequency increases highly predicts the decoupling of the thermal transport across different directions. Furthermore the sensitivity of the in-plane transport increases with the increase in the offset distance between the pump and the probe beams (keeping the spot size constant). From the Fig.~\ref{fig:4}, it is evident that $k_{\perp}$ and $k_{\parallel}$ can not be predicted directly, since both the parameters in the equation occur in the definition of n iterations. However $k_{\perp}$ is described by a more complicated relation by which the two parameters ($\alpha_{\perp}$ and $\alpha_{\parallel}$) are not correlated.

\begin{table}[H]
\centering
\caption{Parameters set constant for fitting and sensitivity analysis}
\label{tab:params_constant}
\begin{tabular}{ll}
\toprule
\textbf{Parameter} & \textbf{Value} \\
\midrule
Layer thickness & \SI{520}{\nano\meter} SL (10 layers) \\
Period thickness & \SI{52}{\nano\meter} (AlAs/GaAs) \\
$R_{\text{12}}$ (\si{\meter\squared\kelvin\per\watt}) & $7.0 \times 10^{-7}$ (from Au/GaAs measurements, $R_\text{th}$) \\
$k_3 = k_{\text{GaAs}}$ (\si{\watt\per\meter\per\kelvin}) & 53.5 \\
$\alpha_3 = \alpha_{\text{GaAs}}$ (\si{\meter\squared\per\second}) & $3.1 \times 10^{-5}$ \\
Input Power & \SI{2.0}{\milli\watt} (probe), \SI{40}{\milli\watt} (pump) \\
Spot Size  & \SI{2.1}{\micro\meter} (pump \& probe) \\
$k_1 = k_{\text{Au}}$ (\si{\watt\per\meter\per\kelvin}) & $105 \pm 9.9$ \\
$\alpha_1 = \alpha_{\text{Au}}$ (\si{\meter\squared\per\second}) & $1.23 \times 10^{-5}$ \\
$l_1 = l_{\text{Au}}$ (\si{\nano\meter}) & 50 \\
\bottomrule
\end{tabular}
\end{table}
In-plane thermal conductivity ($k_{\parallel}$) sensitivity exhibits negative values of $-2.000$ at \SI{10}{\kilo\hertz} under zero beam displacement, transitioning through $-5.000$ at \SI{100}{\kilo\hertz} before reaching $5.000$ at \SI{2}{\mega\hertz}. Concurrently, cross-plane thermal conductivity ($k_{\perp}$) sensitivity initiates at $1.700$ at \SI{10}{\kilo\hertz}, escalates to $7.000$ at \SI{1}{\mega\hertz}, then declines to $-1.000$ at \SI{2}{\mega\hertz}. When beam displacement extends to \SI{1.5}{\micro\meter}, $k_{\parallel}$ sensitivity adjusts to $-1.250$ at \SI{10}{\kilo\hertz}, progresses through $-2.750$ at \SI{100}{\kilo\hertz}, and ultimately reaches $2.250$ at \SI{2}{\mega\hertz}. Corresponding $k_{\perp}$ values under displacement conditions register $0.735$ at \SI{10}{\kilo\hertz}, ascend to $3.650$ at \SI{1}{\mega\hertz}, then diminish to $-0.750$ at \SI{2}{\mega\hertz}.

Spatial sensitivity distributions demonstrate systematic variations with beam displacement magnitude. At zero displacement, $k_{\perp}$ sensitivity measures $0.18$ at \SI{10}{\kilo\hertz}, increasing to $0.35$ at \SI{2}{\mega\hertz}, while $k_{\parallel}$ sensitivity progresses from $0.05$ to $0.15$ across identical frequencies. With \SI{0.10}{\micro\meter} displacement, $k_{\perp}$ values range from $0.15$ to $0.28$, and $k_{\parallel}$ extends from $0.12$ to $0.22$. At \SI{1.5}{\micro\meter} maximum displacement, $k_{\perp}$ sensitivity reduces to $0.01$--$0.02$ across all frequencies, whereas $k_{\parallel}$ sensitivity peaks at $0.48$ at \SI{10}{\kilo\hertz} before diminishing to $0.39$ at \SI{2}{\mega\hertz}. These quantitative relationships elucidate fundamental interdependencies between modulation frequency, spatial configuration, and directional thermal property measurement fidelity.

\section{Global Optimization of Thermal Response}

Simultaneous determination of in-plane and cross-plane thermal properties presents a challenging optimization problem due to parameter coupling and measurement noise. The objective function—relating thermal parameters to measured signals—typically exhibits complex characteristics including non-convexity, high dimensionality, and computational expense. We frame this as a black-box optimization problem where experimental measurements $y_i$ result from an unknown function $\tau$ operating on input conditions $\mathbf{x}_i$: $y_i = \tau(\mathbf{x}_i)$. In our implementation, $\mathbf{x}_i = (r, f)$ represents the experimental variables (beam offset distance $r$ and modulation frequency $f$), while $y_i$ corresponds to the normalized phase and amplitude of the thermoreflectance response.

When working with Gaussian Processes (GPs), we often have a set of $n$ training inputs and $n^*$ test inputs. The covariance relationships between these sets are captured in matrices such as $K(\mathbf{x}, \mathbf{x}^*)$, where each entry represents the covariance between a training and a test point. To obtain predictions, we begin by considering the joint prior distribution over the function values, and then condition this prior on the observed data. Conceptually, this can be thought of as generating many possible functions from the prior and rejecting those that do not fit the data, although in practice, we efficiently achieve this by conditioning the Gaussian distribution.

The posterior distribution over the test function values, $\tau^*$, given the training inputs $\mathbf{x}$, observed outputs $\tau$, and test inputs $\mathbf{x}^*$, is itself Gaussian. Respective mean and covariance are given by \cite{Rasmussen2006}:

\begin{equation}
\tau^* \mid \mathbf{x}, \tau, \mathbf{x}^* \sim \mathcal{N}\left(K(\mathbf{x}^*, \mathbf{x}) K(\mathbf{x}, \mathbf{x})^{-1} \tau, K(\mathbf{x}^*, \mathbf{x}^*) - K(\mathbf{x}^*, \mathbf{x}) K(\mathbf{x}, \mathbf{x})^{-1} K(\mathbf{x}, \mathbf{x}^*)\right)
\label{eq:posterior}
\end{equation}

\paragraph{}This allows direct sampling of predictions from the posterior once the mean and covariance are evaluated. In real-world scenarios, however, observations are often contaminated with noise. Assuming additive independent Gaussian noise with variance $\epsilon^2$, the covariance between noisy observations is adjusted by adding $\epsilon^2$ to the diagonal elements of the covariance matrix. As a result, the joint prior distribution over the noisy training outputs $\mathbf{y}$ and the test function values $\tau^*$ becomes \cite{Rasmussen2006}:

\begin{equation}
\begin{bmatrix}
\tau \\
\tau^*
\end{bmatrix}
= \mathcal{N}\left(
\mathbf{0},
\begin{bmatrix}
K(\mathbf{x}, \mathbf{x}) + \epsilon^2 I & K(\mathbf{x}, \mathbf{x}^*) \\
K(\mathbf{x}^*, \mathbf{x}) & K(\mathbf{x}^*, \mathbf{x}^*)
\end{bmatrix}
\right)
\label{eq:joint_prior}
\end{equation}

Conditioning on the observed noisy outputs yields the key predictive equations for Gaussian Process regression under noisy observations indicating the predictive mean $\mathbb{E}$ and predictive covariance simultaneously \cite{Rasmussen2006}:

\begin{align}
\mathbb{E}[\tau^* \mid \mathbf{x}, \mathbf{y}, \mathbf{x}^*] &= K(\mathbf{x}^*, \mathbf{x}) [K(\mathbf{x}, \mathbf{x}) + \epsilon^2 I]^{-1} \mathbf{y}
\label{eq:predictive_mean} \\
\text{Cov}(\tau^*) &= K(\mathbf{x}^*, \mathbf{x}^*) - K(\mathbf{x}^*, \mathbf{x}) [K(\mathbf{x}, \mathbf{x}) + \epsilon^2 I]^{-1} K(\mathbf{x}, \mathbf{x}^*)
\label{eq:predictive_cov}
\end{align}

These approaches provide a foundation for making predictions with uncertainty, incorporating both the underlying function variability and the observation noise. Extension to multidimensional inputs is straightforward, requiring only appropriate evaluations of the covariance function, although visualization becomes less trivial in higher dimensions.

\subsection{Bayesian Optimisation}

The proposed Bayesian optimization framework leverages Gaussian Process Regression (GPR) to estimate thermal transport properties through a probabilistic approach \cite{Rasmussen2006}. At its core, the method models the experimental response $y_i = g(\mathbf{x}_i)$ as a black-box function, where $\mathbf{x}_i = (r,f)$ represents spatial offset \si{\micro\meter} and modulation frequency, while $y_i$ corresponds to the normalized thermal response. The Gaussian process prior $Y_{1:n} \sim \mathcal{N}(0, \sigma^2I)$ captures the inherent uncertainty in thermal measurements, with the symmetric kernel function $K(\mathbf{x},\dot{\mathbf{x}}) = \frac{1}{2\sigma_{length}^2}\|\mathbf{x}-\dot{\mathbf{x}}\|^2$ governing correlations between data points \cite{Rasmussen2006}.

\begin{figure}[htbp]
\centering
\includegraphics[width=0.85\textwidth]{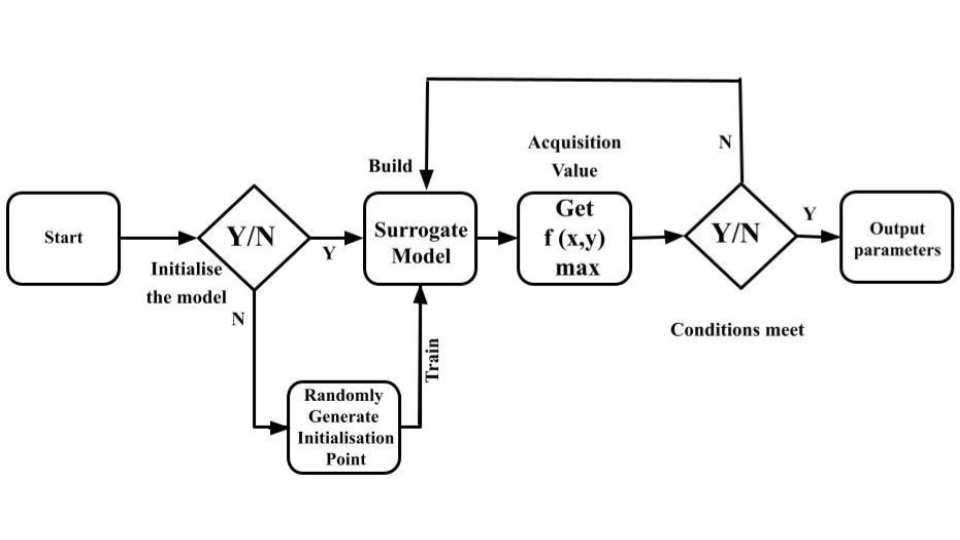}
\includegraphics[width=0.95\textwidth]{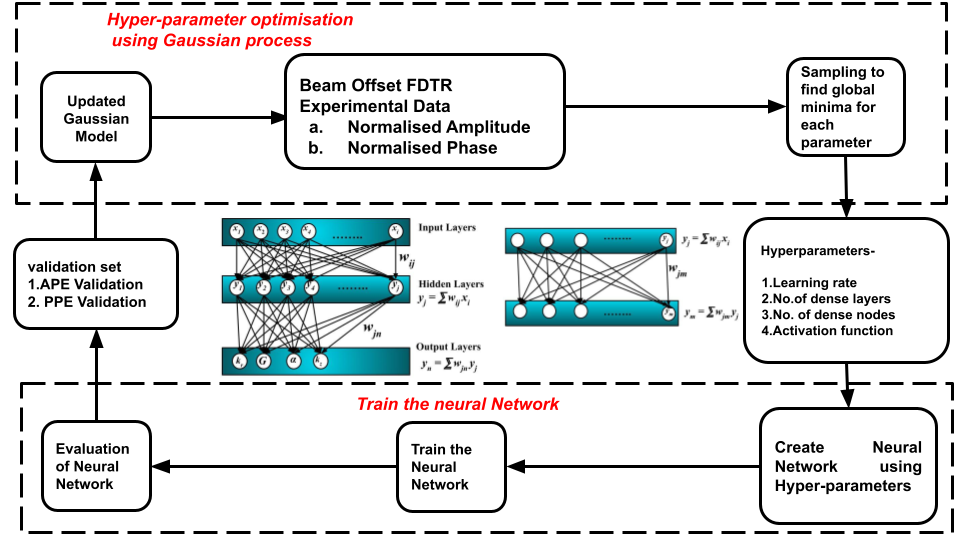}
\caption{Left: Data flow in the model. Right: Bayesian optimization process for amplitude and phase.}
\label{fig:5}
\end{figure}
\paragraph{}A key innovation of this work is the dual estimation strategy for anisotropic thermal properties. The first approach determines in-plane thermal conductivity ($k_{\parallel}$) and diffusivity ($\alpha_{\parallel}$) from thermal boundary resistance ($R_{th}$) using frequency-domain thermoreflectance (FDTR) measurements. Simultaneously, a second method derives cross-plane properties through Bayesian inversion of the same beam-offset data, achieving \SI{95}{\percent} confidence intervals through rigorous uncertainty quantification \cite{Gelman2013}. The joint posterior distribution:

\begin{equation}
\begin{bmatrix}
\tau \\
g
\end{bmatrix}
\sim \mathcal{N}\left(
\begin{bmatrix}
K(\mathbf{x},\mathbf{x}) & K(\mathbf{x},\dot{\mathbf{x}}) \\
K(\dot{\mathbf{x}},\mathbf{x}) & K(\dot{\mathbf{x}},\dot{\mathbf{x}})
\end{bmatrix}
\begin{bmatrix}
\mu \\
\mu'
\end{bmatrix},
\Sigma
\right)
\end{equation}

This facilitates dual estimation of the mean of both in-plane and cross-plane thermal parameters by incorporating both beam-offset ($\mu$) and frequency-modulation ($\mu'$) approaches within a unified probabilistic framework.The covariance matrix $\Sigma$ plays a fundamental role in Gaussian Process regression by quantifying the uncertainty and relationships between data points. This symmetric, positive-definite matrix emerges naturally from the kernel-based formulation of GPs, where it encodes both the variance of individual predictions and the covariance between different points in the input space.

Structurally, $\Sigma$ decomposes into four block matrices:

\begin{equation}
\Sigma = 
\begin{bmatrix}
\Sigma_{\tau\tau} & \Sigma_{\tau g} \\
\Sigma_{g\tau} & \Sigma_{gg}
\end{bmatrix}
\end{equation}

The covariance structure comprises $\Sigma_{\tau\tau} = K(\mathbf{x},\mathbf{x}) + \sigma_n^2\mathbf{I}$ for observed points (including noise variance $\sigma_n^2$), $\Sigma_{gg} = K(\dot{\mathbf{x}},\dot{\mathbf{x}})$ for prediction points, and their interaction $\Sigma_{\tau g} = \Sigma_{g\tau}^\top = K(\mathbf{x},\dot{\mathbf{x}})$ governing how observations inform predictions. The spectral properties of $\Sigma$ fundamentally govern the Gaussian Process's behavior through three principal mechanisms: (1) the diagonal elements $\Sigma_{\tau \tau}$ and $\Sigma_{gg}$ directly determine prediction confidence intervals, scaling the uncertainty bounds at each evaluation point (frequency scans for individual beam offset value); (2) the eigenvalue spectrum's decay rate characterizes the model's effective degrees of freedom, indicating how many independent patterns can be captured; and (3) the matrix condition number $\kappa(\Sigma) = \lambda_{\text{max}}/\lambda_{\text{min}}$ critically affects numerical stability during the Cholesky decomposition required for matrix inversion, with ill-conditioned systems ($\kappa \gg 1$) introducing significant numerical errors in posterior computations.

\FloatBarrier
\begin{table}[H]
\centering
\caption{Bayesian Optimization Framework for Anisotropic Thermal Property Estimation}
\label{tab:bayesian_framework}
\begin{tabular}{p{0.18\textwidth}p{0.38\textwidth}p{0.38\textwidth}}
\toprule
\textbf{Component} & \textbf{Mathematical Formulation} & \textbf{Physical Interpretation} \\
\midrule
\textbf{Input Variables} & 
Experimental measurements: $\mathbf{x}_i = (A_i, \phi_i)$ where:
\begin{itemize}
\item $A_i$: normalized amplitude
\item $\phi_i$: normalised phase (degrees)
\end{itemize} & 
Experimental thermal response data from Beam offset FDTR measurements with anisotropic thermal information \\
**Estimation Strategy** &
Bayesian inversion: $P(k_{\parallel}, k_{\perp}, R_{th} | A, \phi) \propto P(A, \phi | k_{\parallel}, k_{\perp}, R_{th}) P(k_{\parallel}, k_{\perp}, R_{th})$ &
Probabilistic determination of all thermal properties simultaneously from amplitude and phase data using Bayes' theorem \cite{Bui2006} \\

\textbf{Output Parameters} &
Estimated thermal properties:
\begin{itemize}
\item $k_{\parallel}$: in-plane thermal conductivity
\item $k_{\perp}$: cross-plane thermal conductivity  
\item $R_{th}$: thermal boundary resistance
\end{itemize} &
Anisotropic thermal transport properties of the material system extracted through Bayesian inference \\

\textbf{GP Prior} &
$Y_{1:n} \sim \mathcal{N}(0, \sigma^2I)$ &
Probabilistic model representing prior belief about thermal response before observing data \\

\textbf{Kernel Function} &
$K(\mathbf{x},\dot{\mathbf{x}}) = \frac{1}{2\sigma_{length}^2}\|\mathbf{x}-\dot{\mathbf{x}}\|^2$ &
Correlation structure between thermal response measurements based on experimental conditions \\

\textbf{Covariance Matrix} &
$\Sigma = \begin{bmatrix}
\Sigma_{\tau\tau} & \Sigma_{\tau g} \\
\Sigma_{g\tau} & \Sigma_{gg}
\end{bmatrix}$ where:
\begin{itemize}
\item $\Sigma_{\tau\tau} = K(\mathbf{x},\mathbf{x}) + \sigma_n^2\mathbf{I}$
\item $\Sigma_{gg} = K(\dot{\mathbf{x}},\dot{\mathbf{x}})$
\item $\Sigma_{\tau g} = \Sigma_{g\tau}^\top = K(\mathbf{x},\dot{\mathbf{x}})$
\end{itemize} &
Encodes uncertainty in thermal measurements and relationships between different experimental conditions \\

\textbf{Estimation Strategy} &
Bayesian inversion: $P(k_{\parallel}, k_{\perp}, R_{th} | A, \phi) \propto P(A, \phi | k_{\parallel}, k_{\perp}, R_{th}) P(k_{\parallel}, k_{\perp}, R_{th})$ \cite{Bui2006} &
Probabilistic determination of all thermal properties simultaneously from amplitude and phase data using Bayes' theorem \cite{Bui2006} \\

\textbf{Uncertainty Quantification} &
\SI{95}{\percent} credible intervals from posterior distribution of $k_{\parallel}, k_{\perp}, R_{th}$ &
Rigorous statistical bounds on estimated thermal parameters accounting for measurement noise and model uncertainty \\

\textbf{Optimization Process} &
Sequential Bayesian updating: $\mathbf{x}_{n+1} = \arg\max EI(\mathbf{x})$ &
Adaptive experimental design that selects next measurement to maximize information gain about thermal properties \\
\bottomrule
\end{tabular}
\end{table}

Overall, the Bayesian architecture provides several advantages for thermal characterization. First, the adaptive sequential design optimally explores the parameter space, focusing measurements where uncertainty is highest. Second, the length scale parameter $\sigma_{length}$ automatically adjusts individual layer ($M_n$) property correlations based on experimental conditions. Finally, the iterative posterior updating mechanism \cite{MacKay2003} enables continuous refinement of thermal estimates as new data becomes available. This combination of GPR modeling with Bayesian inference establishes a robust framework for anisotropic thermal property determination that surpasses traditional curve-fitting methods in both accuracy and reliability.

\section{Results and Discussion}
Validation begins with the isotropic GaAs substrate, whose thermal conductivity exhibits minimal directional variation. Measurements yield $k_{\parallel}$ and $k_{\perp}$ values differing by only \SI{2.4}{\percent}, confirming the substrate's near-perfect isotropy as documented in Table~\ref{tab:thermal_properties}. To establish method validity, we employ a 2-layer Au/GaAs reference structure, comparing results against the 10-period AlAs/GaAs superlattice under investigation. The GaAs substrate provides a consistent baseline, while the reference sample enables evaluation of thermal boundary resistances in multi-layer systems. The calculation of the thermal parameters with Au/GaAs as the reference sample is described in Eq. (:~\ref{eq:37}) and (~\ref{eq:38}). Experimental measurements show excellent agreement with established literature values \cite{Bui2006}.
\subsection{Validation Using Transversely Isotropic Materials}

The GaAs substrate's isotropic character provides a critical validation benchmark. As shown in Table~\ref{tab:thermal_properties}, measured $k_{\parallel}$ and $k_{\perp}$ values demonstrate the expected equivalence within \SI{2.415}{\percent} uncertainty ~\cite{Chatterjee2024}, confirming both the substrate's isotropic nature and the measurement system's accuracy. 
\begin{table}[H]
\centering
\caption{Thermal Property validation Using Transversely Isotropic Materials}
\label{tab:thermal_properties}
\begin{tabular}{lcccc}
\toprule
\textbf{Symmetry} & \textbf{Sample} & \boldmath$k_{\text{GaAs}}$ \textbf{(\si{\watt\per\meter\per\kelvin})} & \boldmath$R_{12}$ \textbf{($\times 10^{-7}$ \si{\meter\squared\kelvin\per\watt})} \\
\midrule
Cross-plane & Au/GaAs & $53.13 \pm 2.3$ & $8.09\pm 0.22$ \\
In-plane & Au/GaAs & $54.14 \pm 1.74$ (isotropic) & $8.11\pm 0.21$ \\
\bottomrule
\end{tabular}
\end{table}

\begin{figure}[H]
\centering
\includegraphics[width=0.98\textwidth]{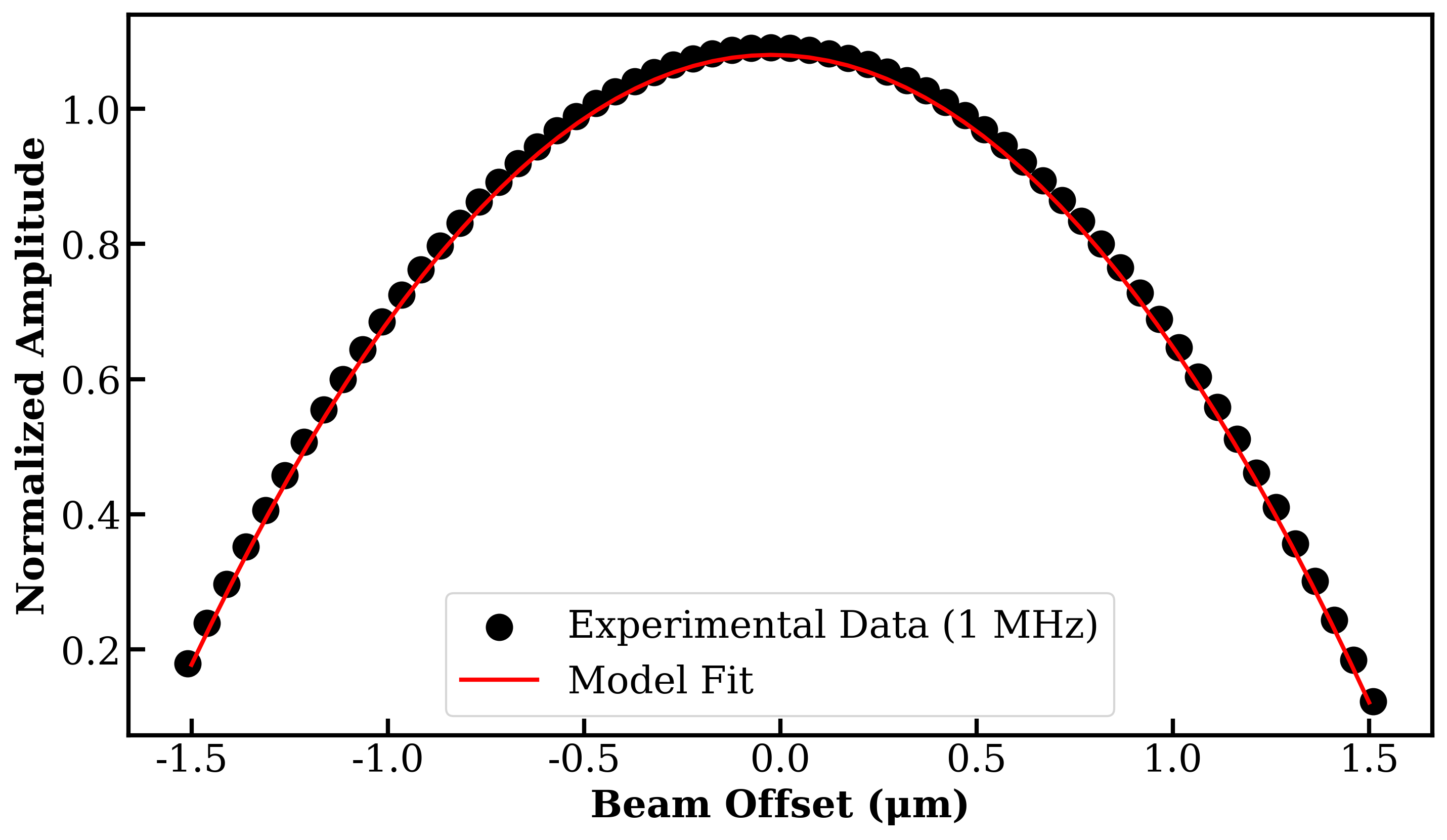}
\includegraphics[width=0.98\textwidth]{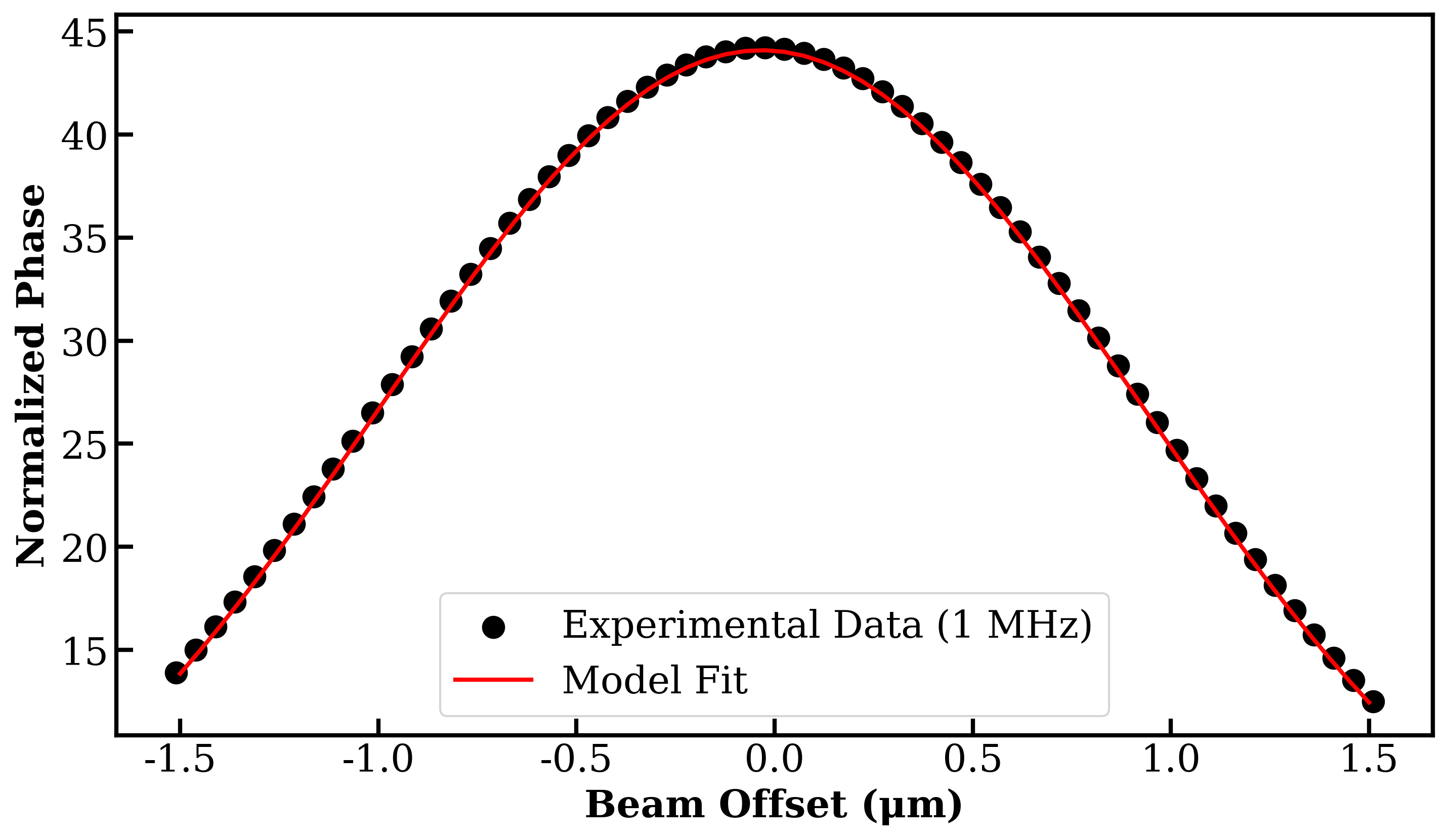}
\caption{Comparison of model introduced in this paper with model from \cite{Pawlak2020}: (a) Normalized Amplitude (b) Normalized Phase; vs offset distance}
\label{fig:7}
\end{figure}
\subsection{Anistropic Thermal Parameter Estimation for AlAs/GaAs Superlattice}
 The anisotropic thermal conductivity and boundary resistance of the AlAs/GaAs nanostructured superlattice were measured at symmetric lateral offsets between the pump and probe beams, scanned across variable ranges of pump modulation frequency. Results from (Table \ref{tab:thermal_properties_selected_offsets}) indicate that the uncertainty in measurement of in-plane thermal conductivity ($k_\parallel$) decreases slightly with larger beam offsets, while the cross-plane value ($k_\perp$ ) exhibits a minor increasing trend.
\begin{table}[H]
\centering
\caption{Anisotropic thermal properties extracted at symmetric beam offset positions using BO-FDTR}
\label{tab:thermal_properties_selected_offsets}
\begin{tabular}{lccc}
\toprule
\textbf{Beam Offset} & \textbf{In-plane Thermal} & \textbf{Cross-plane Thermal} & \textbf{Thermal Boundary} \\
\textbf{(\si{\micro\meter})} & \textbf{$k_{\parallel}$ (\si{\watt\per\meter\per\kelvin})} & \textbf{$k_{\perp}$ (\si{\watt\per\meter\per\kelvin})} & \textbf{$R_{23}$ ($\times 10^{-7}$ \si{\meter\squared\kelvin\per\watt})} \\
\midrule
$\pm 0.10$ & $37.05 \pm 0.32$ & $14.80 \pm 0.05$ & $5.85 \pm 0.02$ \\
$\pm 0.50$ & $37.15 \pm 0.22$ & $14.90 \pm 0.10$ & $5.90 \pm 0.05$ \\
$\pm 1.0$ & $37.55 \pm 0.14$ & $14.65 \pm 0.15$ & $5.97 \pm 0.08$ \\
$\pm 1.5$ & $37.85 \pm 0.05$ & $14.40 \pm 0.20$ & $6.12 \pm 0.12$ \\
\bottomrule
\end{tabular}
\end{table}
The adjoining table (Table \ref{tab:thermal_properties_frequencies}) reports the same uncertainty in properties extracted at zero beam offset but across different modulation frequencies. The data demonstrate that the measured thermal properties remain relatively consistent, showing a relative shift at lower frequency ranges from \SI{10}{\kilo\hertz} to \SI{100}{\kilo\hertz} and minimal variation across the frequency range from \SI{500}{\kilo\hertz} to \SI{2}{\mega\hertz}.
\begin{table}[H]
\centering
\caption{Thermal properties across modulation frequencies at $\Delta r = \SI{0}{\micro\meter}$}
\label{tab:thermal_properties_frequencies}
\begin{tabular}{lccc}
\toprule
\textbf{Modulation} & \textbf{In-plane Thermal} & \textbf{Cross-plane Thermal} & \textbf{Thermal Boundary} \\
\textbf{Frequency} & \textbf{$k_{\parallel}$ (\si{\watt\per\meter\per\kelvin})} & \textbf{$k_{\perp}$ (\si{\watt\per\meter\per\kelvin})} & \textbf{$R_{23}$ ($\times 10^{-7}$ \si{\meter\squared\kelvin\per\watt})} \\
\midrule
\SI{10}{\kilo\hertz} & $36.82 \pm 0.8$ & $15.21 \pm 0.3$ & $6.18 \pm 0.15$ \\
\SI{100}{\kilo\hertz} & $37.15 \pm 0.6$ & $14.98 \pm 0.2$ & $6.12 \pm 0.12$ \\
\SI{500}{\kilo\hertz} & $37.32 \pm 0.5$ & $14.71 \pm 0.15$ & $6.09 \pm 0.10$ \\
\SI{1}{\mega\hertz} & $37.07 \pm 0.17$ & $14.79 \pm 0.2$ & $6.11 \pm 0.13$ \\
\SI{2}{\mega\hertz} & $37.06 \pm 0.29$ & $14.83 \pm 0.4$ & $6.14 \pm 0.18$ \\
\bottomrule
\end{tabular}
\end{table}
The adjoining fitting plot is for the Amplitude and Phase of the 3-D thermoreflectance (TR) model with the thermal parameters shown above in Table \ref{tab:thermal_properties_selected_offsets} and \ref{tab:thermal_properties_frequencies}. Figure (\ref{fig:7}) shows that inclusion of normalised data for amplitude, along with normalised phase, provides a better fitting across the higher frequency of modulation. This suggests better parameter estimation and reduced uncertainty. The parameter for the fitting of both amplitude and phase at different offset values is in Table (~\ref{tab:thermal_properties_frequencies}).
\begin{figure}[H]
\centering
\includegraphics[width=0.85\textwidth]{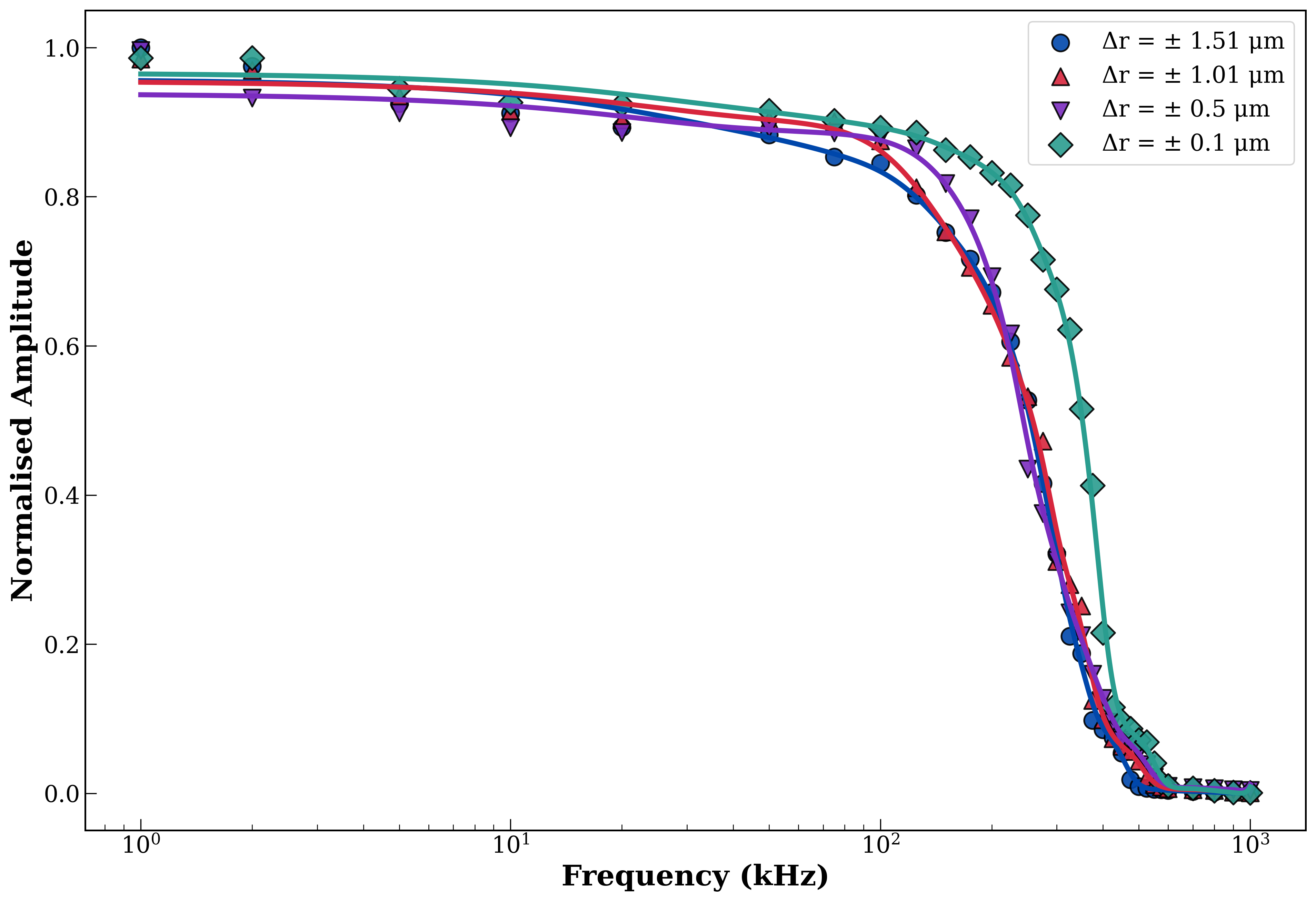}
\includegraphics[width=0.85\textwidth]{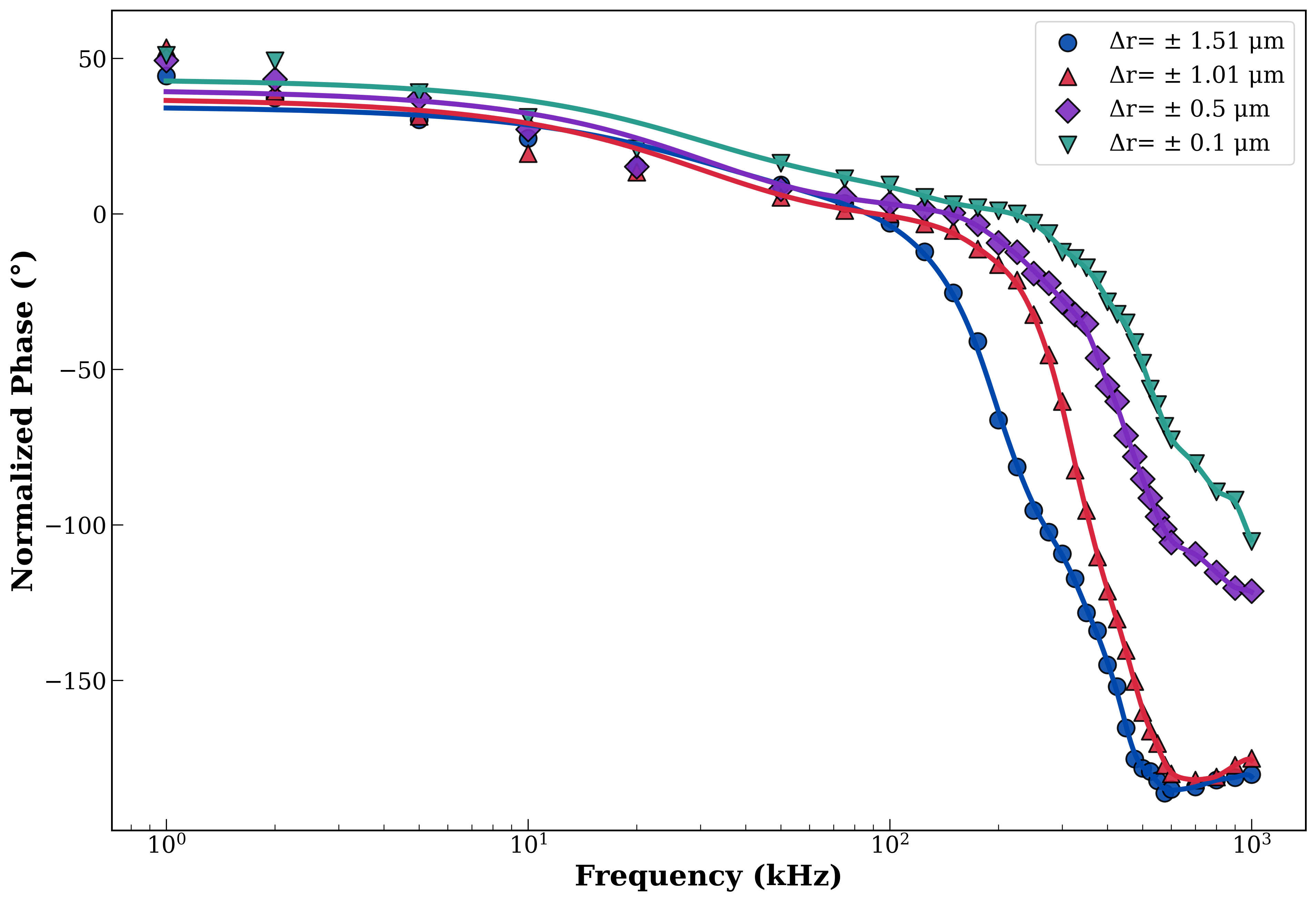}
\caption{Experimental Fit for three-layer system- Normalized (a) Amplitude; (b) Phase; vs modulation Frequency (\si{\kilo\hertz}) (Solid Lines correspond to best fit plots)}
\label{fig:8}
\end{figure}

The normalised Amplitude and Phase for the TR data due to variation of the scanning frequency across different offset postion is shown in Figure (\ref{fig:8}). This diagram validates that as we approach within a particular confidence interval for the pump modulation frequency, uncertainty in determining the anisotropic properties decreases. The parameter for the fitting of both amplitude and phase at different frequency ranges is in Table(~\ref{tab:thermal_properties_selected_offsets}).
\begin{figure}[H]
\centering
\includegraphics[width=0.85\textwidth]{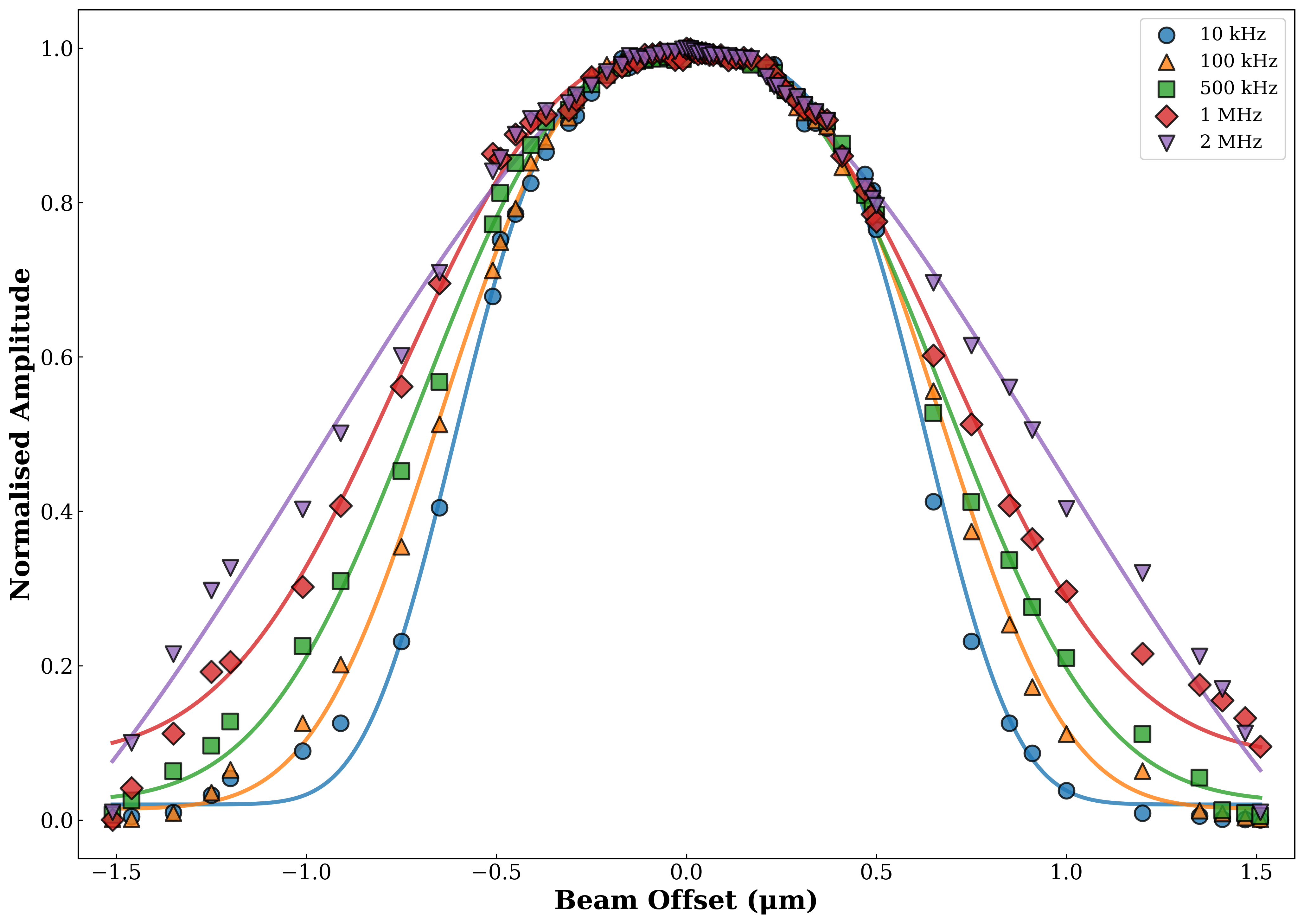}
\includegraphics[width=0.85\textwidth]{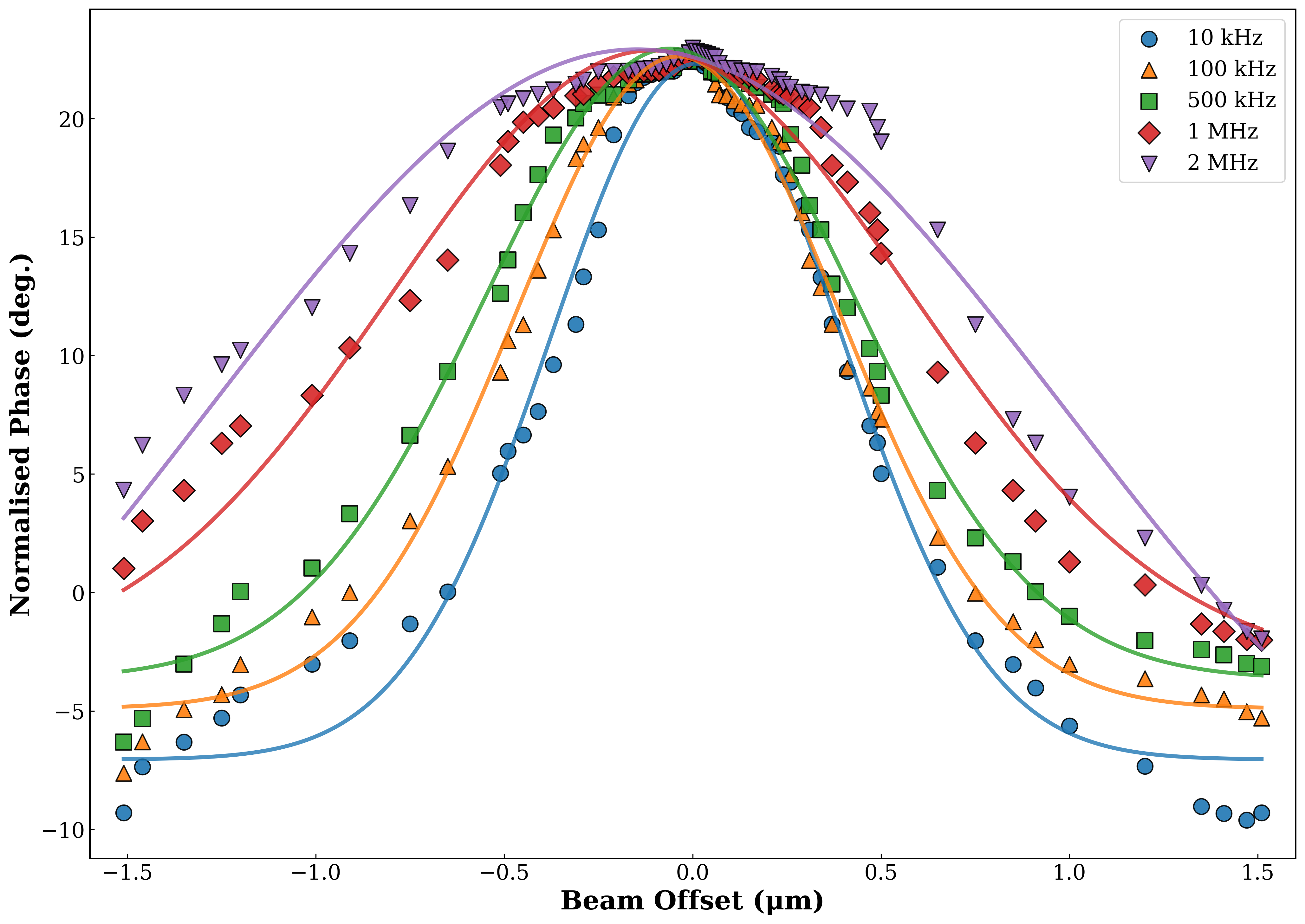}
\caption{Experimental Fit for three-layer system- (a) Normalized Amplitude; (b) Phase; vs Beam offset \si{\micro\meter} (Solid Lines correspond to best fit plots)}
\label{fig:9}
\end{figure}

\begin{table}[H]
\centering
\caption{Thermal property measurements with error estimations (BO-FDTR Approach)}
\label{tab:thermal_properties_fdtr}
\begin{tabular}{lcc}
\toprule
\textbf{Approach} & \boldmath$k_{\text{AlAs/GaAs}}$ \textbf{(\si{\watt\per\meter\per\kelvin})} & \boldmath$R_{23}$ \textbf{($\times 10^{-7}$ \si{\meter\squared\kelvin\per\watt})} \\
\midrule
BO-FDTR ($k_{\perp}$) & $14.90 \pm 0.9$ & $6.13 \pm 0.19$ \\
BO-FDTR ($k_{\parallel}$) & $37.65 \pm 0.25$ & $6.12 \pm 0.06$ \\
FDTR ($k_{\perp}$) & $14.91 \pm 0.24$ & $6.13 \pm 0.04$ \\
FDTR ($k_{\parallel}$) & $36.17 \pm 1.2$ & $6.12 \pm 0.21$ \\
\bottomrule
\end{tabular}
\end{table}

The errors in Table(~\ref{tab:thermal_properties_selected_offsets}) were calculated as the average of the values obtained for each of the frequency scans at constant beam-offset and the difference between the probing and pumping beams (frequency scans). Therefore, the measurement errors relate to the rudimentary errors in determining the thermal parameters from individual frequency scans when the beam offset is also taken as a parameter.

\subsection{Error Analysis and Comparison of Approaches}

The adjoining figure illustrates a comparative analysis of uncertainty in estimating thermal conductivity, diffusivity and thermal boundary resistance using two experimental approaches: frequency scan and beam offset. Each method's results are presented as histograms  in Fig.~\ref{fig:10}, showing the distribution of percentage errors in the each parameter estimation. The frequency scan data is represented by the blue histogram, while the beam offset data is shown in orange. Dashed lines indicate model fits to the respective distributions, highlighting the overall trend of the error spread for each method.

\begin{figure}[H]
\centering
\includegraphics[width=0.62\textwidth]{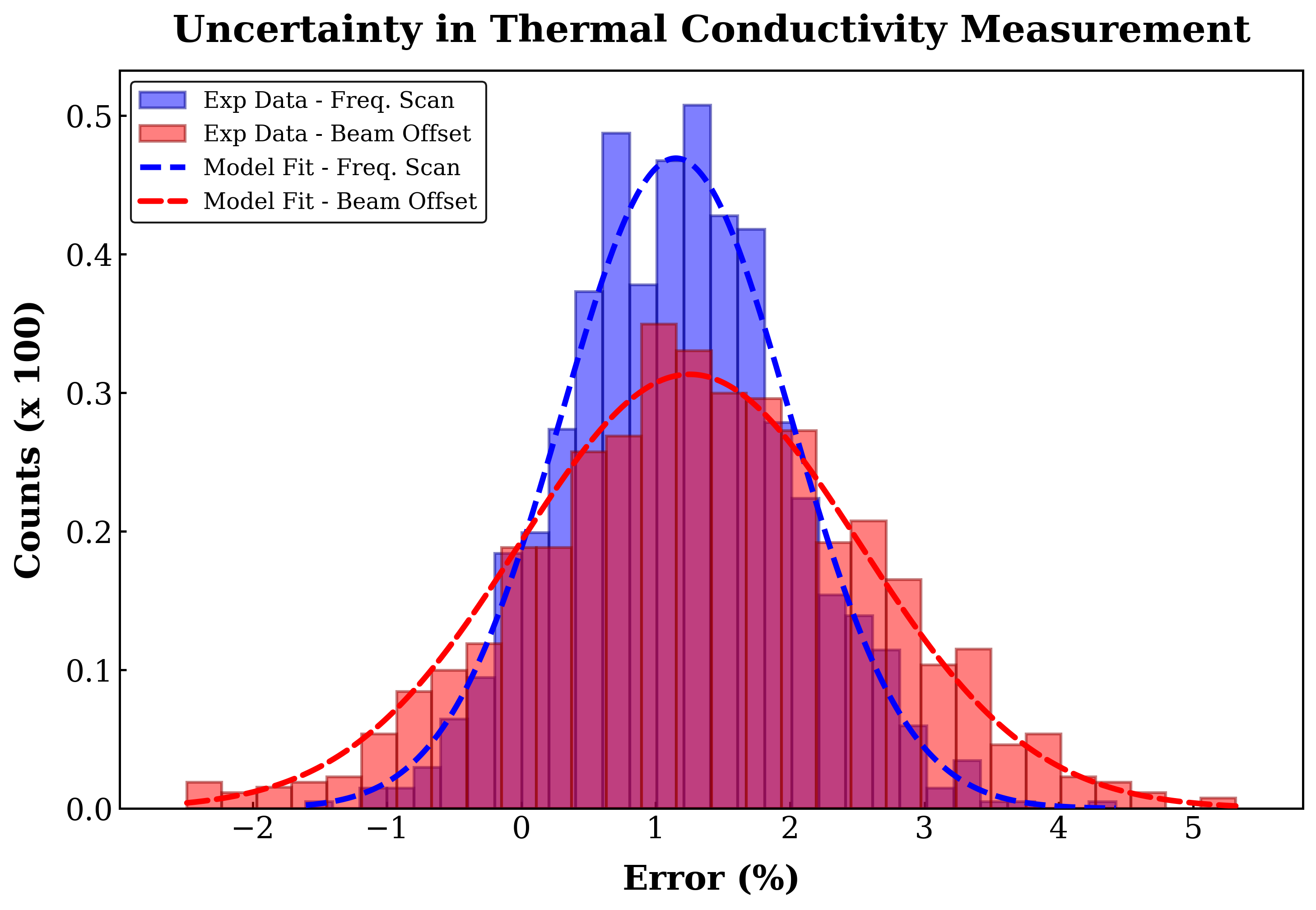}
\includegraphics[width=0.62\textwidth]{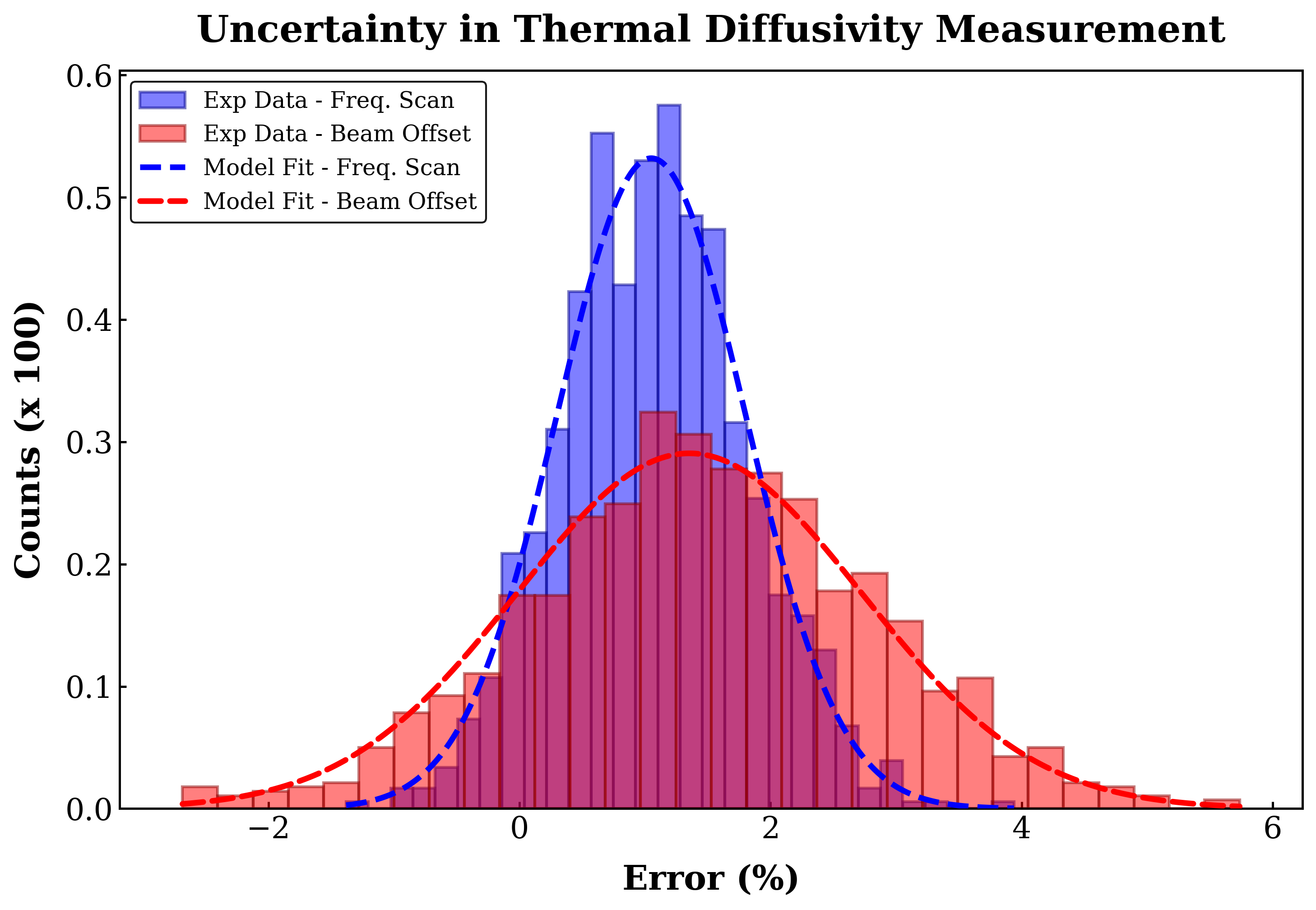}
\includegraphics[width=0.62\textwidth]{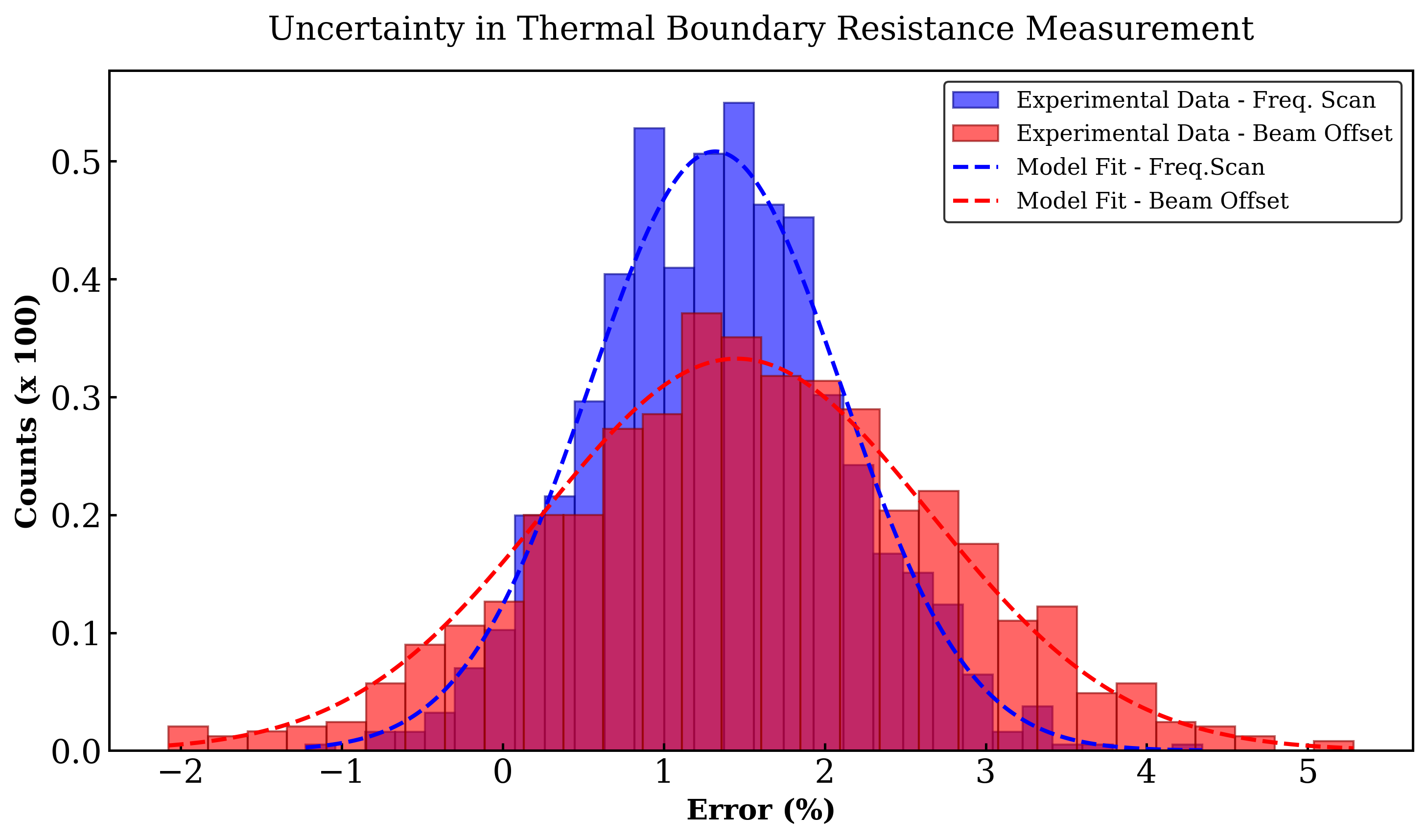}
\caption{Error propagation analysis for three-layer system both input parameters Normalized Amplitude and Phase; vs both approaches (Beam Offset and frequency Scanning) using Bayesian Parameter Estimation algorithm}
\label{fig:10}
\end{figure}

A key observation from the figure is that the frequency scan histogram is significantly narrower than the beam offset histogram. This narrower distribution indicates that the frequency scan method yields more precise results, with the estimation errors more tightly clustered around the mean. It suggests lower variability and greater reliability in the thermal boundary resistance measurements obtained through frequency scanning. In contrast, the wider distribution seen in the beam offset method implies greater uncertainty and variability, likely due to increased sensitivity to experimental noise or limitations in model fitting. Overall, the figure demonstrates that the frequency scan technique provides more consistent and stable estimates of thermal boundary resistance compared to the beam offset approach. After discussions in Section~3, a comparative study between both approaches (Beam offset and Frequency scanning) after initial iterations from the prescribed algorithm is discussed in Fig.~\ref{fig:10}.

\subsection{Implementation of the GPR-BO Framework}

From the perspective of implementing the GPR-BO framework, the best-fitted results for the determination of the thermal parameters has been determined. The key point of discussion from Table VI is the simultaneous and ultra-precise measurement of thermal parameters when more than one parameter can be controlled. The determination is clearly mutually exclusive but changes with the thermal response evaluated from the introduction of beam-offset to the pre-existing FDTR method. Fig.~\ref{fig:10} clearly shows that the applied algorithm predicts the experimental data very well, and is supported by a clear indication that the estimation is accurate for ranges around the zero position.

The in-plane thermal conductivity ($k_{\parallel}$) exhibits a pronounced dependence on beam offset distance, increasing systematically from $37.05 \pm 0.32\,\si{\watt\per\meter\per\kelvin}$ at minimal separation ($\pm 0.10\,\si{\micro\meter}$) to $37.85 \pm 0.05\,\si{\watt\per\meter\per\kelvin}$ at maximum offset ($\pm 1.5\,\si{\micro\meter}$). This progressive enhancement, representing a $2.2\%$ increase in measured value, directly correlates with the improved sensitivity to lateral heat diffusion at larger probe beam displacements. The physical origin of this behavior lies in the increased contribution of in-plane phonon transport pathways when the thermal detection occurs further from the excitation source. Concurrently, the reduction of uncertainty margins from $\pm 0.32\,\si{\watt\per\meter\per\kelvin}$ to $\pm 0.05\,\si{\watt\per\meter\per\kelvin}$ reflects the enhanced signal-to-noise ratio and improved measurement precision achieved at larger beam offsets in this configuration.

\subsection{Systematic Investigation of Anisotropic Thermal Transport}

The systematic investigation of anisotropic thermal transport through symmetric beam offset measurements, as detailed in Table~\ref{tab:thermal_properties_selected_offsets}, reveals fundamental insights into the depth-dependent thermal response of the AlAs/GaAs superlattice structure. The implementation of symmetric offset pairs ($\pm\Delta r$) serves as an internal validation mechanism, ensuring measurement reliability while providing comprehensive characterization of thermal anisotropy.

\begin{figure}[H]
\centering
\includegraphics[width=1.05\textwidth]{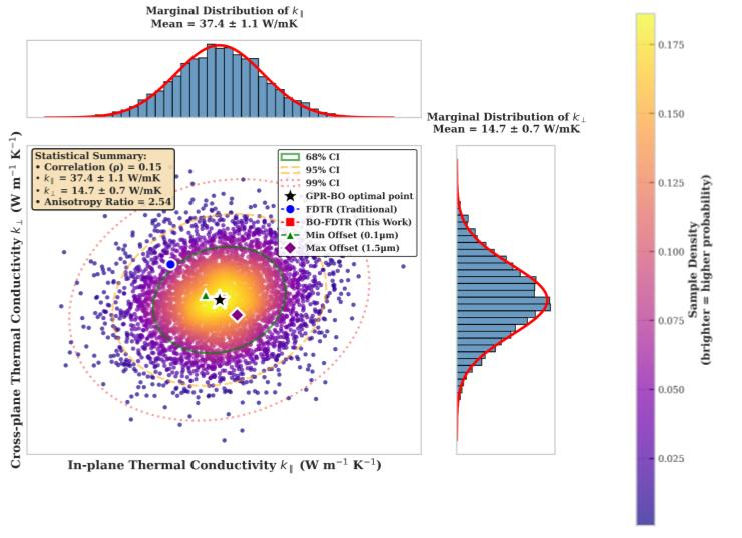}
\caption{Correlation Showing decoupling parameter optimization for the amplitude and phase using beam-offset.}
\label{fig:11}
\end{figure}

The elliptical space defines the search space for in-plane and cross-plane transport, with the weak positive correlation ($\rho=0.15$) being a good sign for independent parameter identification as shown in Fig.~\ref{fig:11}. For the uniformity of the algorithm flow during this parameter estimation analysis, both amplitude and phase have been used as the input datasets for this Bayesian optimization framework. The beam offset TR data squeezes this ellipse horizontally, providing a tight constraint on the possible value of $k_{\parallel}$. The model must simultaneously fit all the other data points in the table and the phase curves from Fig.~\ref{fig:3}. This overlapping set of constraints from multiple experimental conditions ensures the final fit for $k_{\parallel}$ and $k_{\perp}$ is unique and physically meaningful. In essence, the parameter space is no longer a vast, uncertain plane. It is precisely triangulated by adding beam offset to the traditional FDTR signal. Furthermore, frequency scans, each optimally sensitive to one parameter or the other, ultimately converge on a single, best-fit pair of values within a small, confident region. This is the hallmark of robust and valid anisotropic thermal characterization in superlattice sample.

\section{CONCLUSION}

This work presents a beam-offset thermoreflectance methodology that successfully decouples anisotropic thermal transport in semiconductor superlattices. By introducing controlled spatial separation between pump and probe beams, we achieve enhanced sensitivity to in-plane thermal conductivity ($k_{\parallel}$) while maintaining accurate cross-plane ($k_{\perp}$) characterization. The beam-offset configuration fundamentally alters the measurement's sensitivity profile—trading some cross-plane sensitivity for substantially improved in-plane resolution. Physically, this spatial separation enables direct probing of lateral heat diffusion, providing unique access to in-plane transport mechanisms that remain inaccessible to conventional co-located pump-probe configurations. The integration of amplitude and phase measurements within a Bayesian optimization framework enables simultaneous determination of $k_{\parallel}$ and $k_{\perp}$ with high precision. Our analysis demonstrates that combining frequency sweeps with multiple beam offset positions yields optimal accuracy, surpassing conventional single-variable approaches. This methodology's strength lies in its multi-parameter optimization across the complete experimental space (frequency × offset), which treats thermal anisotropy as an explicit fitting parameter. This approach effectively minimizes parameter cross-talk while providing robust confidence intervals for extracted thermal properties.The methodology directly addresses the core challenge of measuring anisotropic thermal conductivity.

\end{document}